\documentclass{aa}

\usepackage{graphicx}
\usepackage{subcaption}
\usepackage{txfonts}
\bibpunct{(}{)}{;}{a}{}{,}
\usepackage{natbib}

\usepackage[colorlinks=true, linkcolor=blue,citecolor=blue]{hyperref}

\begin{document} 

\title{Detection capabilities of the Athena X-IFU for the warm-hot intergalactic medium using gamma-ray burst X-ray afterglows}

\author{Sarah Walsh\inst{1}
        \and Sheila~McBreen\inst{1}
        \and Antonio~Martin-Carrillo\inst{1}
        \and Thomas~Dauser\inst{2}
        \and Nastasha~Wijers\inst{3}
        \and J\"{o}rn~Wilms\inst{2}
        \and Joop~Schaye\inst{3}
        \and Didier~Barret\inst{4}
          }
          
\institute{School of Physics, University College Dublin, Dublin 4, Ireland. \email{sarah.walsh.2@ucdconnect.ie}
\and Remeis Observatory \& ECAP, Universit\"{a}t Erlangen-N\"{u}rnberg, Sternwartstr. 7, 96049 Bamberg, Germany
\and Leiden Observatory, Leiden University, PO Box 9513, NL-2300 RA Leiden, the Netherlands
\and Universit\'{e} de Toulouse, CNRS, Institut de Recherche en Astrophysique et Plan\'{e}tologie, 9 Avenue du colonel Roche, BP 44346, 31028 Toulouse Cedex 4, France
}

\titlerunning{Detection of the WHIM using GRB afterglows with the Athena X-IFU}
\authorrunning{S. Walsh et al.}

\date{Received 19 February 2020 / Accepted 14 July 2020}

\abstract{
At low redshifts, the observed baryonic density falls far short of the total number of baryons predicted. Cosmological simulations suggest that these baryons reside in filamentary gas structures, known as the warm-hot intergalactic medium (WHIM). As a result of the high temperatures of these filaments, the matter is highly ionised such that it absorbs and emits far-UV and soft X-ray photons. \textit{Athena}, the proposed European Space Agency X-ray observatory, aims to detect the `missing' baryons in the WHIM up to redshifts of $z$\,=\,1 through absorption in active galactic nuclei and gamma-ray burst afterglow spectra, allowing for the study of the evolution of these large-scale structures of the Universe.
This work simulates WHIM filaments in the spectra of GRB X-ray afterglows with \textit{Athena} using the SImulation of X-ray TElescopes (SIXTE) framework. 
We investigate the feasibility of their detection with the X-IFU instrument, through \ion{O}{vii} ($E$\,=\,573\,eV) and \ion{O}{viii} ($E$\,=\,674\,eV) absorption features, for a range of equivalent widths imprinted onto GRB afterglow spectra of observed starting fluxes ranging between 10$^{-12}$ and 10$^{-10}$\,erg\,cm$^{-2}$\,s$^{-1}$, in the 0.3--10\,keV energy band.
The analyses of X-IFU spectra by blind line search show that \textit{Athena} will be able to detect \ion{O}{vii}-\ion{O}{viii} absorption pairs with EW\textsubscript{\ion{O}{vii}}\,>\,0.13\,eV and EW\textsubscript{\ion{O}{viii}}\,>\,0.09\,eV for afterglows with \textit{F}\,>\,2\,$\times$\,10$^{-11}$\,erg\,cm$^{-2}$\,s$^{-1}$. This allows for the detection of $\approx$\,45--137 \ion{O}{vii}-\ion{O}{viii} absorbers during the four-year mission lifetime.
The work shows that to obtain an \ion{O}{vii}-\ion{O}{viii} detection of high statistical significance, the local hydrogen column density should be limited at $N_\mathrm{H}$\,<\,8\,$\times$\,10$^{20}$\,cm$^{-2}$.
}
            
\maketitle
   
\section{Introduction}
Predictions from Big Bang nucleosynthesis \citep[BBN;][and references therein]{2016RvMP...88a5004C} combined with observations of the abundance of light elements \citep{burles98}, and the temperature fluctuations of the cosmic microwave background (CMB) taken independently \citep{2003ApJS..148..175S, 2014A&A...571A...1P}, show that baryonic matter accounts for only $\approx$\,4\% of the total cosmic energy content. 
However, at the current epoch, the observed baryon density appears to be much lower \citep{1998ApJ...503..518F}, with studies showing only $\approx$\,60\% of these baryons are detected \citep{2012ApJ...759...23S}. 

Cosmological hydrodynamic simulations, based on the $\Lambda$CDM model, predict that these `missing' baryons reside in gaseous filamentary structures known as the warm-hot intergalactic medium \citep[WHIM;][]{cen99, dave01, cen06}. These baryons are shock heated to temperatures of $10^{5}\,\mathrm{K}<T<10^{7}\,\mathrm{K}$ as baryons in the diffuse intergalactic medium (IGM) accelerate towards sites of structure formation under the growing influence of gravitationally generated potential wells. 
Simulations predict that the WHIM accounts for approximately 30--60\% of the baryonic mass at $z$\,=\,0 \citep{dave01, 2012MNRAS.425.1640T}, making it the largest constituent of the IGM. 

The high temperatures of these filaments result in the matter being highly ionised so that it absorbs and emits far-UV and soft X-ray photons. 
However, these WHIM filaments are of low column density and, therefore, produce a low signal intensity which makes them difficult to detect with the current instrumentation. 
In spite of this, the search for the missing baryons has continued and detections of extragalactic \ion{O}{vii} and \ion{O}{viii} have been claimed \citep{2005ApJ...629..700N} but they are often disputed \citep{2006ApJ...652..189K, 2006ApJ...642L..95W, 2007ApJ...656..129R}, of a single-line or low statistical significance \citep{2010ApJ...715..854N, 2016MNRAS.457.4236B}, misidentifications of the WHIM \citep{2016MNRAS.458L.123N}, or they are local to the background source \citep{2019ApJ...884L..31J}. 
More recently, \citet{2018Natur.558..406N}, later updated by \citet{2018arXiv181103498N}, reported the detection of two \ion{O}{vii} absorption features at redshifts $z$\,=\,0.3551 and $z$\,=\,0.4339 along the sight-line towards blazar \mbox{1ES 1553+113} after a 1.75\,Ms observation with the \textit{XMM-Newton} reflection grating spectrometer. 
\citet{2019ApJ...884L..31J} conducted additional studies towards \mbox{1ES 1553+113}, suggesting that the feature located at $z$\,=\,0.4339 may not originate from the WHIM and is associated, rather, with the local environment of the blazar. 
A study by \citet{2019ApJ...872...83K} quotes a 3.3$\sigma$ detection of \ion{O}{vii} towards \mbox{H1821+643} from a stacked spectrum of \textit{Chandra} observations with an 8\,Ms total exposure.
\citet{2020A&A...634A.106A} performed spectral analysis of \textit{Chandra} and \textit{XMM-Newtown} observations at previously determined FUV redshifts, yielding two X-ray line candidates of \ion{Ne}{ix} \element{He}$\alpha$ and \ion{O}{viii} Ly$\alpha$ at a combined significance of 3.9$\sigma$.
In spite of the large observational efforts, only these few marginal detections have been achieved so far. 
The small equivalent widths of $\approx$ 0.07--0.42\,eV \citep{2009ApJ...697..328B, 2018Natur.558..406N} from WHIM absorption features mean that high signal-to-noise ratio observations are required for their detection. 
This calls for an observatory with a large effective area, high energy resolution, and a low energy threshold in the soft X-ray energy band.

Selected as part of ESA's Cosmic Vision programme, \textit{Athena} will study the evolution of large scale structure through the detection of WHIM filaments to trace the missing baryons in the local universe.  
\textit{Athena} aims to measure the local cosmological baryon density in the WHIM to better than 10\% and to constrain structure formation models in the low-density regime by measuring the redshift distribution and the physical parameters of 200 filaments against bright background sources. 
To achieve this, \textit{Athena} will detect 200 filaments in the WHIM through absorption, 100 towards active galactic nuclei (AGN), and 100 towards bright gamma-ray burst (GRB) afterglows, up to redshifts of $z$\,=\,1  \citep{2013arXiv1306.2324K}. 
The WHIM can be detected by observing the absorption of highly ionised elements such as C, N, O, Ne, and possibly Fe (e.g. \ion{O}{vii}, \ion{O}{viii}, \ion{Ne}{ix}, \ion{Fe}{xvii}). 
The strongest lines expected correspond to the H-like and He-like oxygen ions of the \ion{O}{vii} 1s--2p X-ray resonance line (574\,eV) and the unresolved \ion{O}{viii} 1s--2p X-ray doublet (653.5\,eV, 653.7\,eV). 
A bright X-ray background source such as an AGN \citep[e.g.][]{2017AARv..25....2P} or a GRB \citep[e.g.][]{2017SSRv..207...63W} can be used to detect the WHIM by producing absorption features in its energy spectra \citep{1998ApJ...503L.135P, 1998ApJ...509...56H, 2000ApJ...544L...7F}. 
These sources provide a high probability of detection for the WHIM because they are sufficiently bright and distant to obtain a large statistical sample of lines. 
Bright AGN are common but are typically nearby, having an average redshift of $z$\,$\approx$\,0.8 for flat-spectrum radio quasars (FRSQ) and $z$\,$\approx$\,0.3 for BL-Lacs \citep{2011ApJ...743..171A} and so, they probe relatively short lines of sight. 
Gamma-ray bursts occur at an average redshift of $z$\,$\approx$\,2 \citep{2009MNRAS.397.1177E} and have been detected out to redshifts of 9.4 \citep{2011ApJ...736....7C}. 
This allows for the probing of long lines of sight and can potentially provide multi-filament detections in a single observation. 
In addition, GRBs occur at an approximate rate of 1 GRB per day, with \textit{Fermi}-GBM detecting $\approx$\,250 per year with a $\approx$\,70\% sky coverage \citep{2016ApJS..223...28N}.
Both sources provide similar fluences in the 0.3--10\,keV energy range, but GRBs can provide this fluence in much shorter integration times.
The challenge associated with the use of GRBs as background sources is their transient nature, emitting a considerable percentage of the soft X-ray photons within the first hour of their afterglow phase. 
Therefore, an instrument capable of having a high efficiency to react on target of opportunity (ToO) events is required to probe the WHIM with GRBs.
In spite of this, GRBs should allow Athena to perform its science mission of tracing the missing baryons in GRB afterglow spectra throughout its four-year mission lifetime.

To achieve the goals set out by its science case, \textit{Athena} will have two scientific instruments, the Wide Field Imager \citep[WFI;][]{2017SPIE10397E..0VM} and the X-ray Integral Field Unit \citep[X-IFU;][]{2018SPIE10699E..1GB}. 
The WFI is a silicon-based detector using depleted field effect transistor (DEPFET) active pixel sensor technology that provides imaging in the 0.2--15\,keV energy band over a wide field, simultaneously with spectrally- and time-resolved photon counting. 
The X-IFU is an array of transition edge sensors (TES), which provides spatially high resolution spectroscopy over an energy range of 0.2--12\,keV. 
The instrument has a field of view of $5'$ equivalent diameter and will deliver X-ray spectra with a spectral resolution of 2.5\,eV up to 7\,keV. 
The high spectral resolution of the X-IFU makes it the prime instrument for the detection of the WHIM. 
Both instruments alternately share a moveable mirror system based on silicon pore optics with a focal length of 12\,m and a large effective area of approximately 1.4\,m$^{2}$ at an energy of 1\,keV. 
The baseline configuration of the mirror consists of 15 rows of silicon pore optics. 
The \textit{Athena} mirror assembly offers a defocusing capability, allowing for observations of the brightest X-ray sources of the sky, up to Crab-like intensities, by spreading the telescope point spread function over hundreds of pixels.

This work uses GRB afterglows as background sources for the detection of WHIM filaments. 
Gamma-ray bursts are the most luminous events known to occur in our Universe, consisting of energetic bursts of $\gamma$-rays from deep space. 
Early observations showed that the distribution of GRB duration was bimodal leading to the identification of short and long duration GRBs \citep[e.g.][]{1984Natur.308..434N, 1993ApJ...413L.101K}, where short GRBs have a typical duration (defined as T$_{90}$, the time over which 90\,\% of the total energy release in $\gamma$-rays is recorded) of $T_{90}$\,<\,2\,s while long GRBs have duration $T_{90}$\,>\,2\,s. 
Broad-band observations show that the two duration classes roughly correspond to two types of progenitor systems: long GRBs are related to deaths of massive stars \citep[e.g.][]{1999ApJ...524..262M}, while short GRBs are related to compact binary mergers \citep[e.g.][]{1984SvAL...10..177B,1986ApJ...308L..43P}.

Gamma-ray bursts consist of a prompt phase during which they can emit up to 10$^{53}$ erg over a timescale of seconds and an afterglow phase which can last for days after the prompt phase and emits in energy bands from X-ray to radio. 
Studies of the GRB X-ray afterglow are performed with observatories such as the Neils Gehrels Swift Observatory \citep{2004ApJ...611.1005G}, XMM-Newton \citep{2001A&A...365L...1J}, the Chandra X-ray Observatory \citep{2002PASP..114....1W}, and INTEGRAL \citep{2003A&A...411L...1W}, with their instruments regularly observing GRB sources tens of seconds after the trigger.
The afterglow light curve of most GRBs can be described by a canonical light curve which includes five main temporal components each following a power-law decay of various decay indices \citep[e.g.][]{2006ApJ...642..354Z, 2006ApJ...642..389N, 2006ApJ...647.1213O, 2009MNRAS.397.1177E}.

Initial simulations of the WHIM detectability using GRB afterglows as a background source were implemented and detections were proved to be feasible using the 2\,m$^{2}$ configuration of the \textit{Athena} mirror \citep{2016SPIE.9905E..5FB, brand_thesis}. 
This paper investigates the feasibility of detecting these missing baryons but with the current baseline (1.4\,m$^{2}$ at 1\,keV) mirror configuration. 
It advances the work of \citet{2016SPIE.9905E..5FB} to determine the ability of the X-IFU to recover the characteristics of the WHIM absorption features and to determine the effect of Galactic absorption on the WHIM detection. 
It uses the current \textit{Swift}-XRT GRB afterglow population and the results from EAGLE simulations \citep{2015MNRAS.446..521S, 2015MNRAS.450.1937C} to determine the number of \ion{O}{vii}-\ion{O}{viii} absorption systems that may be detected by the X-IFU during its mission lifetime.

This paper is set out as follows. 
Section \ref{sec:methods} describes the modelling of the GRB X-ray afterglow imprinted with WHIM absorption features and the simulations executed to create mock \textit{Athena} X-IFU spectra. 
Section \ref{sec:data_analysis} discusses the data analysis method to search for WHIM absorbers in the simulated spectra. 
In Sect.~\ref{sec:35mm}, the probability of detecting the WHIM with the \textit{Athena} X-IFU is presented for the defocused configuration of the mirror. 
Section \ref{sec:equiv_widths} reports the equivalent width distributions that are obtained from the simulations and the ability of the X-IFU to reproduce the characteristics of the WHIM. 
The effect of the local hydrogen column density on the detection of WHIM absorption features at soft X-ray energies is investigated in Sect.~\ref{sec:nh_effect}.
The results are discussed in Sect.~\ref{sec:discussion} to assess the ability of \textit{Athena} to discover the missing baryons of the WHIM. 
Finally, the findings of the paper are summarised in Sect.~\ref{sec:conclusions}.

\begin{table*}
\centering
\caption{Spectral parameters used to model the GRB afterglow and WHIM filaments in all simulations}
\begin{tabular}{lllc}
\hline\hline
\textbf{Model Component} & \textbf{Model Parameter} & \textbf{Value}    & \textbf{References} \\ \hline
Local Galaxy    & Galactic Absorption  	& 2 $\times$ 10$^{20}$ cm$^{-2}$ 	&  \\ \hline
GRB Afterglow	& Redshift, $z$\textsubscript{GRB}  & 2.0 & 1 \\
			    & Intrinsic Absorption 	& 1\,$\times$\,10$^{22}$\,cm$^{-2}$   	& 1 \\
				& Photon Index         	& 2.0                   	& 1 \\ \hline
\ion{O}{vii} Line at 574 eV	    & Redshift, $z$\textsubscript{WHIM}      & 0.0--0.5     & \\
                        & Equivalent Width, EW\textsubscript{\ion{O}{vii}} & 0.07--0.42 eV          & 2, 3 \\
                        & $\sigma$      & 0.1 eV ($b$\,=\,73\,km/s, $T\approx5\times10^{6}$\,K)                    & 4  \\ \hline
\ion{O}{viii} Line at 654 eV (2 line only)	& Redshift, $z$\textsubscript{WHIM}  & 0.0--0.5  &           \\
                        & Equivalent Width EW\textsubscript{\ion{O}{viii}} & 2/3 of \ion{O}{vii} & 4 \\
                        & $\sigma$        & 0.1 eV ($b$\,=\,65\,km/s, $T\approx4\times10^{6}$\,K)    & 4\\ \hline
\end{tabular}
\label{tab:model_components}
\tablebib{(1) \citet{2009MNRAS.397.1177E}; (2) \citet{2009ApJ...697..328B};
(3) \citet{2018Natur.558..406N}; (4) \citet{2016SPIE.9905E..5FB}.
}
\end{table*}

\section{Simulating GRB afterglows} \label{sec:methods}
Specifically designed for the \textit{Athena} X-IFU, \textit{xifupipeline} executes the simulations of GRB spectra in the SImulation of X-ray TElescopes (SIXTE) framework\footnote{https://www.sternwarte.uni-erlangen.de/research/sixte/} \citep{2019A&A...630A..66D}. 
The simulations use the current 1.4\,m$^{2}$ (at 1\,keV) baseline configuration\footnote{Area response file: sixte\_xifu\_cc\_baselineconf\_20180821.arf} of the \textit{Athena} mirror.
As the mirror offers a defocusing capability to maximise the photons collected by the X-IFU instrument during the observation of bright sources, the simulations are performed with the 35\,mm defocused configuration.
The main inputs for the pipeline are the SIMulation inPUT file \citep{schmid13} which describes the light curve and spectrum of the observed source, the XML file which describes the X-IFU calorimeter, and the advanced XML which describes the grading scheme for detected photons. 
User-defined parameters are also input into the pipeline such as the pointing of the telescope in right ascension ($\alpha$) and declination ($\Delta$), the exposure time, and whether or not crosstalk between pixels is considered. 

In all simulations, the exposure time was set to 50\,ks, in accordance with the average exposure time allocated to the X-IFU for a given observation. 
Results for different exposure times can be obtained by scaling the flux of the results presented in this paper.
Crosstalk was simulated between pixels and the grading of photons accepted was those of only high resolution where $\Delta$E\,$\leq$\,2.5\,eV. 

The SIMPUT file contained one source modelled as a GRB afterglow by the input spectrum and light curve. 
The spectrum of the GRB afterglow was modelled by an absorbed power law comprising three main parts; an absorption function describing the Galactic absorption from the Milky Way galaxy, modelled by \texttt{tbabs} \citep{2000ApJ...542..914W}; an absorption function describing the intrinsic absorption from the host galaxy of the GRB, modelled by \texttt{ztbabs}; and, a power-law, \texttt{pegpwrlw}, describing the GRB afterglow emission. 
The WHIM was modelled by two faint Gaussian absorption lines, \texttt{zgauss}, which were redshifted from the rest energy of the lines. 
The most prominent line expected is \ion{O}{vii} and is located at a rest-frame energy of 574\,eV, while the second line representing \ion{O}{viii} has a rest-frame energy of 654\,eV. 
Both lines are assumed to originate from the same cosmic structure, and so, share the same redshift, $z$\textsubscript{WHIM}. 
The parameters adopted for modelling the GRB afterglow spectrum and WHIM filaments are shown in Table \ref{tab:model_components}.

A decaying power-law models the light curve of the GRB afterglow, which follows a relation between flux and time of $F$\,$\propto$\,$t^{\alpha}$, where $\alpha$\,$=$\,--1.2 is the power-law index \citep{2013MNRAS.428..729M}, in the relevant time range between two and 30 hours after the initial outburst occurred.
The starting flux \textit{F}\textsubscript{start} of the GRB is defined as the observed flux of the GRB afterglow, measured in the 0.3--10\,keV energy range, when the observation begins. 
This was varied from 10$^{-12}$ to 10$^{-10}$ erg cm$^{-2}$ s$^{-1}$ in accordance with fluxes of observed GRBs by the \textit{Swift} X-ray observatory four hours after the initial burst.
Located at varying redshifts, six WHIM lines of different equivalent widths from the expected range of the WHIM were simulated for each flux. 
The redshift was varied randomly between $z$\textsubscript{WHIM}\,=\,0.0 and 0.5 to accommodate the search for local baryons within the \textit{Athena} X-IFU energy band, while the equivalent widths were varied in the range EW\textsubscript{\ion{O}{vii}}\,=\,0.07--0.42\,eV \citep{2009ApJ...697..328B, 2018Natur.558..406N}.
For simplicity, the ratio between \ion{O}{vii} and \ion{O}{viii} remained fixed for all input spectra and is given by the ratio of the ion oscillator strengths so that EW\textsubscript{\ion{O}{viii}}\,=\,2/3\,$\cdot$\,EW\textsubscript{\ion{O}{vii}}.
Fifty-one observations were simulated for each combination of equivalent width and GRB starting flux, with the redshift of the WHIM filament varying randomly in the given range.
This provides a significant sample with a reasonable computation time.

The detection probability of two absorption features of Gaussian form can also be derived from Poissonian statistics. 
However, to verify the simulation method and the analysis method, and to create a simulation environment that incorporates the detector response, source light curves, and source spectra, the Monte Carlo simulations are performed to analyse the detectability of \ion{O}{vii} and \ion{O}{viii} absorption features in GRB afterglow spectra. 
A comparison of both methods in shown in Appendix A.

\begin{figure*}
\centering
\resizebox{\hsize}{!}
{
\begin{subfigure}{.5\textwidth}
\centering
\includegraphics[width=\linewidth]{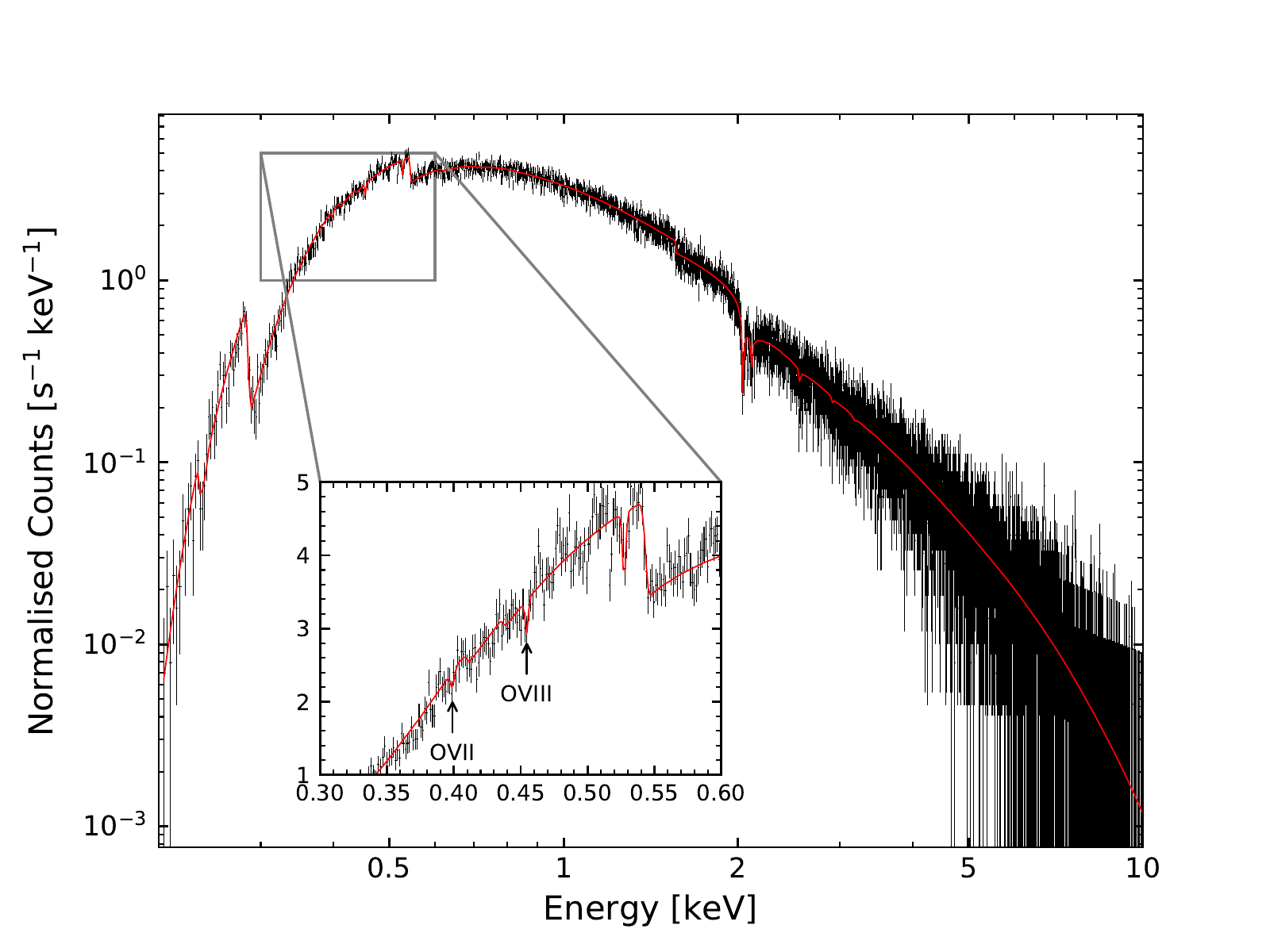}
\label{fig:spectrum_5e-12}
\end{subfigure}%
\hfill
\begin{subfigure}{.5\textwidth}
\centering
\includegraphics[width=\linewidth]{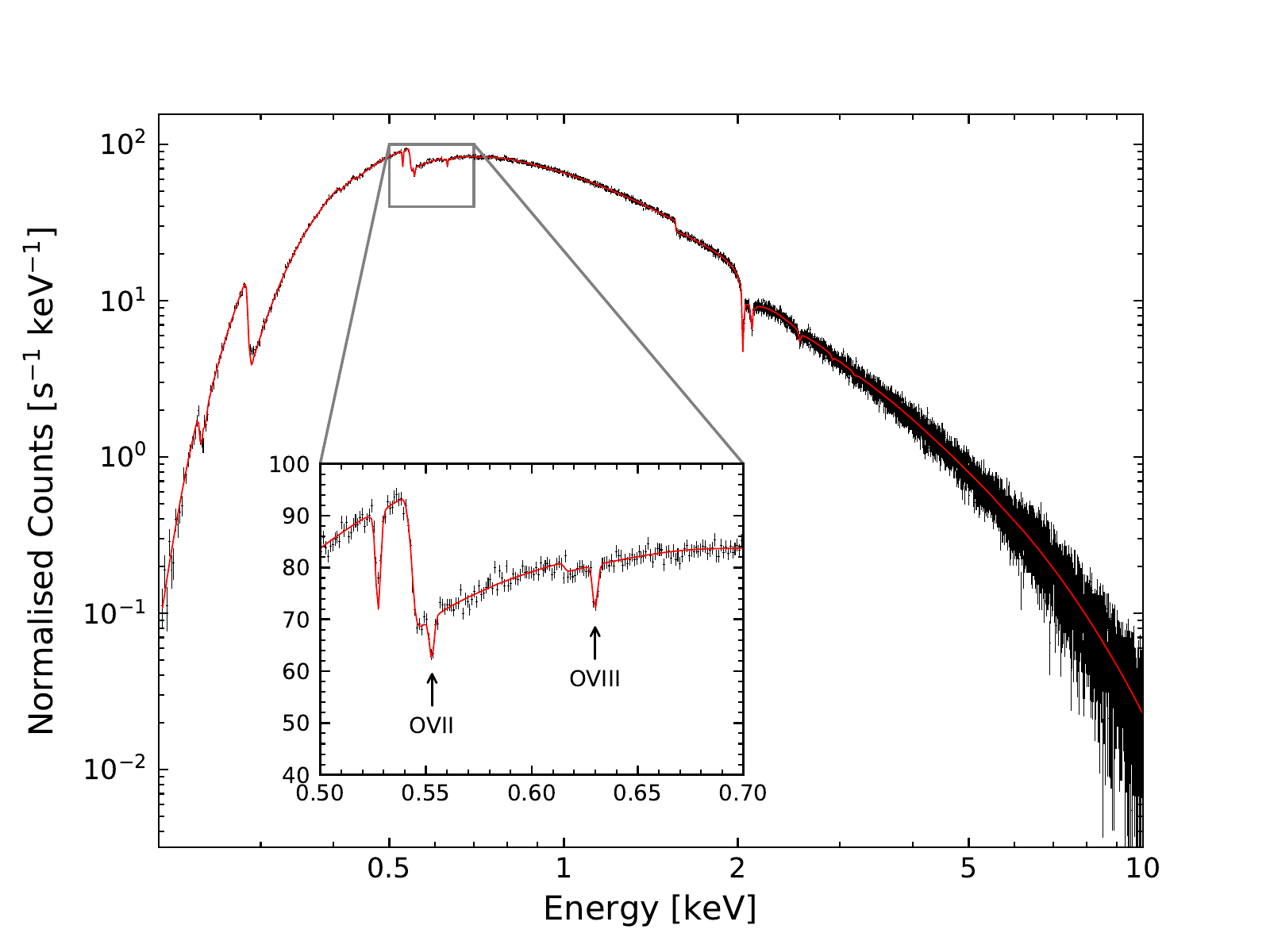}
\label{fig:spectrum1e-10}
\end{subfigure}%
}
\caption{50\,ks simulated X-IFU energy spectra of a GRB afterglow ($z$\textsubscript{GRB}\,=\,2) imprinted by \ion{O}{vii} (EW\textsubscript{\ion{O}{vii}} = 0.35\,eV) and \ion{O}{viii} (EW\textsubscript{\ion{O}{viii}} = 0.23\,eV) WHIM absorbers. Both spectra are binned using the optimal binning scheme of \citet{2016A&A...587A.151K}. \textit{Left:} GRB absorbed starting flux 5\,$\times$\,10$^{-12}$\,erg\,cm$^{-1}$\,s$^{-1}$ and $z$\textsubscript{WHIM}\,=\,0.4388. \textit{Right:} GRB absorbed starting flux 1\,$\times$\,10$^{-10}$\,erg\,cm$^{-1}$\,s$^{-1}$ and $z$\textsubscript{WHIM}\,=\,0.0382. The insets show the energy range around the WHIM absorption features. Note: The absorption feature located at $\approx$\,0.53\,eV is the O resonance line of Galactic origin and is contained within the \texttt{tbabs} model component.}
\label{fig:xifu_spectra}
\end{figure*}

\section{Data analysis} \label{sec:data_analysis}
\subsection{Spectral fitting} \label{sec:spectral_fitting}
The simulated spectra (e.g. Fig~\ref{fig:xifu_spectra}) were analysed in XSPEC v12.10 \citep{arnaud96}. 
The optimal binning scheme of \citet{2016A&A...587A.151K} with ftools \texttt{ftgrouppha} and the Cash statistic $\mathcal{C}$ \citep{1979ApJ...228..939C, 2017A&A...605A..51K} were used for the grouping and fitting of the spectral data. 

The spectra were fitted in the energy range of 0.3--10 keV, which is within the energy range of the X-IFU. 
The model used for fitting the data was the same model that was used to generate the event data; an intrinsically absorbed power law, at the redshift of the GRB, with a Galactic absorption component and Gaussian absorption lines to represent the WHIM filaments. 
The absorption components were modelled with the \texttt{tbabs} and \texttt{ztbabs} components in XSPEC while \texttt{wilm} abundances \citep{2000ApJ...542..914W} and \texttt{vern} cross-sections \citep{1996ApJ...465..487V} were used. 
The redshift of the GRB, and Galactic absorption were assumed to be known and were fixed during the fitting process, while all other parameters were free to vary. 
The WHIM lines were fitted by a blind line search as with real data the properties of the intergalactic gas and the distance at which it is located will not be known. 
The redshift range in which the blind line fitting process searches for absorption features was set to 0.0\,$\leq$\,$z$\textsubscript{WHIM}\,$\leq$\,0.5. 
The continuum parameters were initialised with the input model parameters while the energies of the absorption lines were set to their rest energies.
The blind line search began at a redshift of $z$\textsubscript{WHIM}\,=\,0. 
At each iteration of the search the redshift was increased by $\Delta z$\textsubscript{WHIM}\,=\,0.002 and the corresponding fit statistic was determined. 
This was repeated for all redshift values in the search range, producing a list of fit statistic values, one for each redshift trial.
From this list the redshift value of the best-fit model was determined by calculating the minimum $\mathcal{C}$-value. 
This redshift value was taken to be the most significant line positions of the \ion{O}{vii} and \ion{O}{viii} absorption features in the spectrum.

To assess the significance of these features, Monte Carlo simulations were performed, as described in \citet{2002ApJ...571..545P}, to eliminate the possibility that the absorption features were chance statistical fluctuations in the data. 
The best-fit model for the spectrum with no absorption features, that is, the continuum only\footnote{\texttt{tbabs}$\cdot$\texttt{ztbabs}$\cdot$\texttt{pegpwrlw}}, was assumed as the null hypothesis model and was simulated 100 times using XSPEC's \texttt{fakeit} command.
Each simulated spectrum was then fit with the null hypothesis model to obtain the fit statistic.
Two Gaussian absorption features, separated by a fixed energy, were simultaneously searched for using the same blind line fitting process as before in the same redshift range, producing a list of fit statistic values for each redshift trial in a simulated spectrum. 
The minimum $\mathcal{C}$-value from the blind line search and the $\mathcal{C}$-value from the continuum only fit for each of the simulated spectra were recorded.
A distribution of the $\Delta\mathcal{C}$-values was obtained, where $\Delta\mathcal{C}$ is the difference between the fit statistic of the continuum fit and the minimum fit statistic of the blind line search. 
The $\Delta\mathcal{C}$-values obtained from each simulated spectra create a distribution for which fitting statistical fluctuations in the spectrum improves the statistical fit of the continuum model. 
From this distribution, the $\Delta\mathcal{C}$-value that corresponds to a 90$\%$ significant detection for the original lines found in the data was determined. 
An original line is found to be significant if its measured $\Delta\mathcal{C}$-value is higher than the 90$\%$ significant $\Delta\mathcal{C}$-value from the distribution of the simulated set of spectra. 
Finally, it must be determined whether the features found by the significant detections are from the input WHIM lines. 
This was achieved by defining an additional margin of half the spectral energy resolution of the X-IFU, $\Delta z$\textsubscript{WHIM}\,=\,1\,$\times$\,10$^{-3}$ or $\approx$1.25\,eV, on the energy of the detected absorption features relative to the known redshifted energy of the input lines. 
If an absorption feature is detected within the defined margin of their input energy, it is classed as a correct detection, otherwise, it is a false-alarm detection resulting from noise that has been fitted in the spectrum.

\subsection{Detection probabilities of absorption features}
All simulated observations of a given $F$\textsubscript{start} and line equivalent width are combined to calculate the line detection probability by
\begin{equation} \label{detection_probability}
P(\text{\small line found, \textit{F}\textsubscript{start}, EW\normalsize}) = \dfrac{N(\text{\small sig. detection, correct $z$\textsubscript{WHIM}, \textit{F}\textsubscript{start}, EW\normalsize})}{N(\text{\small \textit{F}\textsubscript{start}, EW\normalsize})},
\end{equation}
where $N$(\textit{F}\textsubscript{start}, EW) is the total number of simulated observations for a given absorbed starting flux, \textit{F}\textsubscript{start} and equivalent width, EW and $N$(significant detection, correct, $z$\textsubscript{WHIM}, \textit{F}\textsubscript{start}, EW) is the total number of significant detections of $N$(\textit{F}\textsubscript{start}, EW) that were found at the correct WHIM redshift \citep{2016SPIE.9905E..5FB}. 
Equation \ref{detection_probability} provides the probability that a line or \ion{O}{vii}-\ion{O}{viii} line pair is detected with a minimum significance of 90\% and that the redshift it was detected at is within the energy margin defined in Sect.~\ref{sec:spectral_fitting}. 
The probability function can be used to determine the minimum absorbed starting flux of a GRB afterglow or the minimum number of spectral counts required to detect the absorption line or line pair of given strengths.

In addition to the probability of a line being detected, the false-alarm probability can be calculated.
The false-alarm detection probability is the probability that an absorption feature has been determined to be significant to a minimum of a 90\% threshold but lies outside the tolerated energy margin of the true value.
It is determined by
\begin{equation} \label{false_alarm_probability}
P(\text{\small false-alarm, \textit{F}\textsubscript{start}, EW}\normalsize) = \dfrac{N(\text{\small sig. detection, incorrect $z$\textsubscript{WHIM}, \textit{F}\textsubscript{start}, EW}\normalsize)}{N(\text{\small sig. detection, \textit{F}\textsubscript{start}, EW} \normalsize)}.
\end{equation}
Equation \ref{false_alarm_probability} gives an indication of the occurrence rate of selecting a noise feature within the spectra to be significant, over a real absorption feature, during the spectral analysis.
The false-alarm probability is an important quantity because it could lead to a misinterpretation and over-estimate of the WHIM abundance, if not taken into account. 

\section{Results} \label{sec:results}
\subsection{Detection of \ion{O}{vii} and \ion{O}{viii} in X-IFU spectra}
\label{sec:35mm}
\begin{figure}
\centering
\begin{subfigure}{.49\textwidth}
\centering
\includegraphics[width=1.\linewidth]{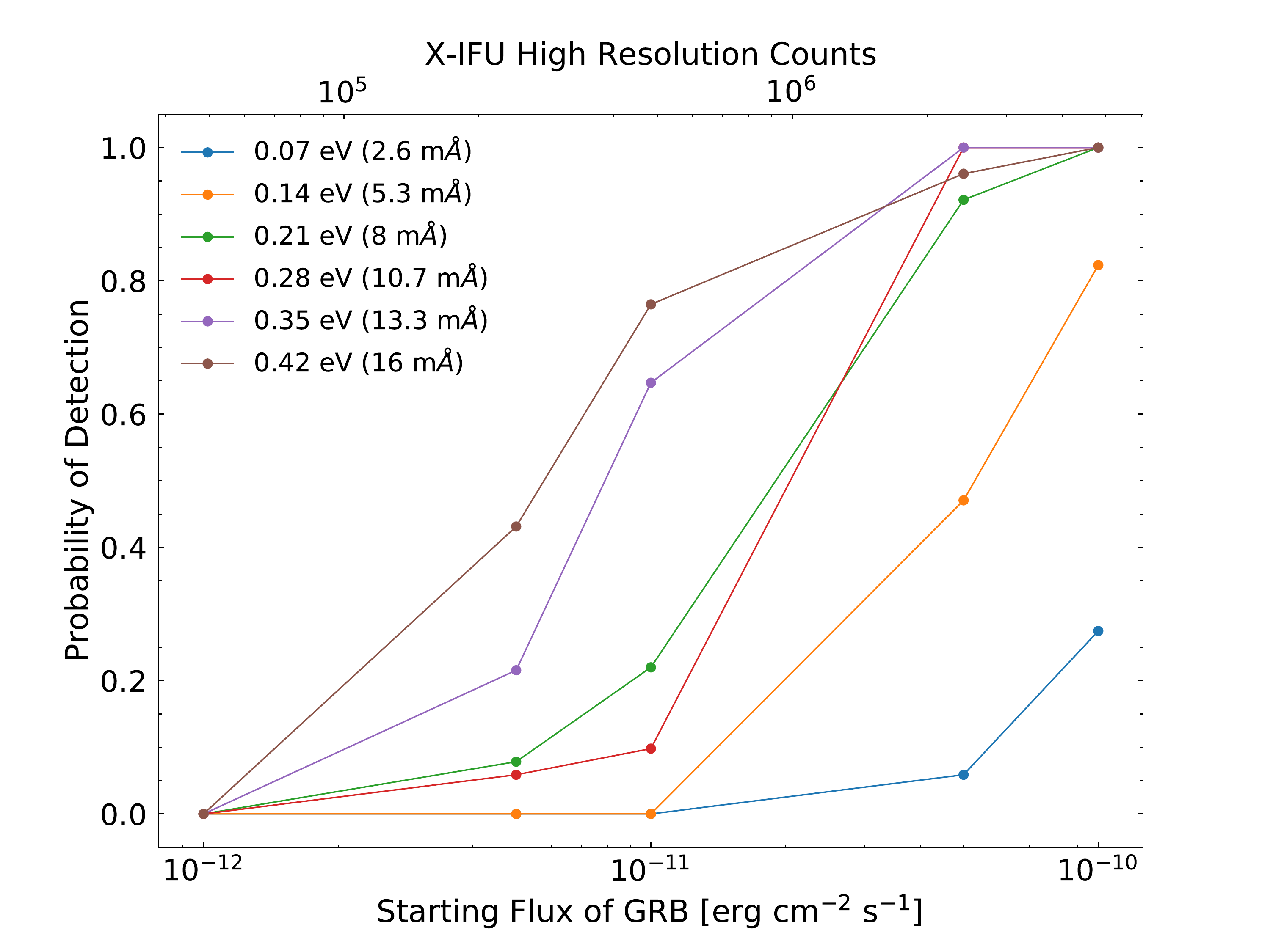}
\end{subfigure}%
\hfill
\begin{subfigure}{.49\textwidth}
\centering
\includegraphics[width=1.\linewidth]{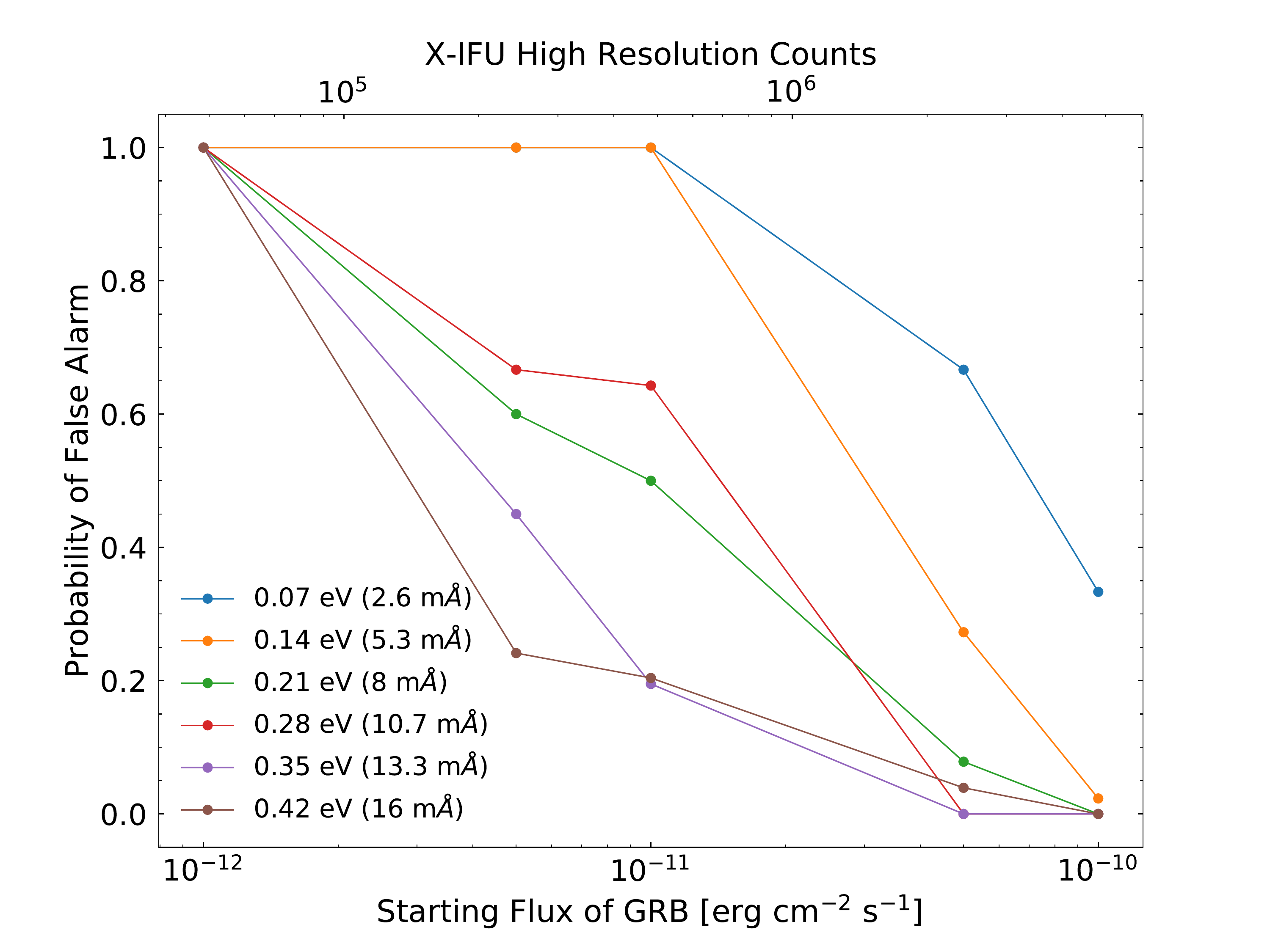}
\end{subfigure}%
\caption{\textit{Top:} Detection probability by \ion{O}{vii} and \ion{O}{viii} absorption features for each EW simulated in the 35\,mm defocused configuration of the \textit{Athena} mirror. \textit{Bottom:} False-alarm probability of WHIM detection by \ion{O}{vii} and \ion{O}{viii} absorption features for each EW simulated EW in the 35\,mm defocused configuration of the \textit{Athena} mirror. EWs listed are those of \ion{O}{vii}.}
\label{fig:35mm_detect}
\end{figure}

Gamma-ray burst afterglow spectra containing a WHIM imprint of \ion{O}{vii} and \ion{O}{viii} were simulated with the mirror in a 35\,mm defocused configuration and were analysed using the method in Sect.~\ref{sec:data_analysis}.
The results are shown in Fig.~\ref{fig:35mm_detect}. 
The top panel shows the detection probability for observed starting fluxes of GRB afterglows in the range 10$^{-12}$--10$^{-10}$\,erg\,cm$^{-2}$\,s$^{-1}$ in the 0.3--10\,keV energy band and for line equivalent widths of EW\textsubscript{\ion{O}{vii}}\,=\,0.07--0.42\,eV, with \ion{O}{viii} having an equivalent width of 2/3\,$\cdot$\,EW\textsubscript{\ion{O}{vii}}. 

The probability of detecting the WHIM by \ion{O}{vii}-\ion{O}{viii} paired absorption features with at least a 90\% significance increases with increasing afterglow flux and with increasing line equivalent widths. 
For the brightest GRB afterglow simulated of \textit{F}\textsubscript{start}\,=\,10$^{-10}$\,erg\,cm$^{-2}$\,s$^{-1}$, there is at least an 80$\%$ chance of detecting an \ion{O}{vii}-\ion{O}{viii} pair with EW\textsubscript{\ion{O}{vii}}\,>\,0.14\,eV.
The weakest lines simulated with EW\textsubscript{\ion{O}{vii}}\,=\,0.07\,eV provide a $\sim$ 30$\%$ chance of detection for the same starting flux. 
However, at the lowest afterglow flux of \textit{F}\textsubscript{start}\,=\,10$^{-12}$\,erg\,cm$^{-2}$\,s$^{-1}$, none of the simulated lines are detected.

To obtain a 90\% probability of detecting an \ion{O}{vii}-\ion{O}{viii} absorption pair with at least a 90\% significance, \textit{Athena} should target GRB afterglows that have \textit{F}\textsubscript{start}\,>\,3\,$\times$\,10$^{-11}$\,erg\,cm$^{-2}$\,s$^{-1}$. 
This limit allows the detection of absorbers with EW\textsubscript{\ion{O}{vii}}\,>\,0.2\,eV. 
To obtain a 50\% probability of detecting an \ion{O}{vii}-\ion{O}{viii} absorption feature with a minimum of 90\% significance, an afterglow of \textit{F}\textsubscript{start}\,$=$\,6--20\,$\times$\,10$^{-12}$\,erg\,cm$^{-2}$\,s$^{-1}$ must be observed, limiting lines of EW\textsubscript{\ion{O}{vii}}\,>\,0.3\,eV.
A starting flux of 5\,$\times$\,10$^{-11}$\,erg\,cm$^{-2}$\,s$^{-1}$ will be required from a GRB afterglow to have a 50\% probability of detecting lines EW\textsubscript{\ion{O}{vii}}\,>\,0.15\,eV. 
The weakest absorber of 0.07\,eV requires \textit{F}\textsubscript{start}\,$=$\,8\,$\times$\,10$^{-11}$\,erg\,cm$^{-2}$\,s$^{-1}$ during a 50\,ks exposure to obtain a 20\% probability of detecting absorption features with a 90\% significance.
These starting flux and equivalent width limits for 90\%, 50\%, and 20\% detection probabilities are listed for reference in Table \ref{tab:whim_detection}.

The bottom panel of Fig.~\ref{fig:35mm_detect} shows the probability that a significant detection of two absorption features separated by the same fixed energy difference is made at an energy that does not match the input energy (within $\Delta$E\,=\,1.25\,eV).
This results in a false-alarm detection of an \ion{O}{vii}-\ion{O}{viii} pair.
The probability of a false-alarm detection for a given afterglow flux and line equivalent width is calculated by Equation \ref{false_alarm_probability}.
Opposite to the detection probability, the false-alarm probability decreases with increasing afterglow flux and with increasing line equivalent width.
For afterglow spectra with starting fluxes of \textit{F}\,=\,10$^{-12}$\,erg\,cm$^{-2}$\,s$^{-1}$ the chance of a misdetection is near 100$\%$ for all equivalent widths. 
Lines with EW\textsubscript{\ion{O}{vii}}\,$\geq$\,0.21\,eV see the false-alarm probability drop to less than 5$\%$ when the starting flux of the GRB afterglow is \textit{F}\,=\,5\,$\times$\,10$^{-11}$\,erg\,cm$^{-2}$\,s$^{-1}$, while lines of EW\textsubscript{\ion{O}{vii}}\,=\,0.14\,eV reach this level at \textit{F}\,=\,\,10$^{-10}$\,erg\,cm$^{-2}$\,s$^{-1}$.

\begin{table}
\centering
\caption{Minimum \ion{O}{vii} observed equivalent width that can be detected at 90\%, 50\%, and 20\% detection probabilities for different GRB afterglow starting fluxes during a 50\,ks observation.}
\begin{tabular}{ccrrr}
\hline\hline
\multicolumn{1}{p{1.8cm}}{\centering $F$\textsubscript{start,\,0.3--10\,keV}  \\ {\footnotesize[erg\,cm$^{-2}$\,s$^{-1}$]}}
&\multicolumn{1}{p{1.7cm}}{\centering 
$F$\textsubscript{0.3--10\,keV}  \\ {\footnotesize[erg\,cm$^{-2}$\,s$^{-1}$]}}
&\multicolumn{1}{p{1cm}}{\centering \small EW\textsubscript{90\%} \\ {\footnotesize[eV]}}
&\multicolumn{1}{p{1cm}}{\centering \small EW\textsubscript{50\%} \\ {\footnotesize[eV]}}
&\multicolumn{1}{p{1cm}}{\centering \small EW\textsubscript{20\%} \\ {\footnotesize[eV]}} \\ \hline
1\,$\times$\,10$^{-10}$  &   3.3\,$\times$\,10$^{-11}$ & 0.17 & 0.10 & <\,0.07 \\
5\,$\times$\,10$^{-11}$  &   1.7\,$\times$\,10$^{-11}$ & 0.20 & 0.14 & 0.09 \\
1\,$\times$\,10$^{-11}$  &   3.3\,$\times$\,10$^{-12}$ & >\,0.42 & 0.33 & 0.23 \\
5\,$\times$\,10$^{-12}$  &   1.7\,$\times$\,10$^{-12}$ & >\,0.42 & >\,0.42 & 0.34 \\
1\,$\times$\,10$^{-12}$  &   3.3\,$\times$\,10$^{-13}$ & >\,0.42 & >\,0.42 & >\,0.42 \\ \hline
\end{tabular}
\label{tab:whim_detection}
\tablefoot{Column 1: simulated starting flux (erg\,cm$^{-2}$\,s$^{-1}$) of the GRB afterglow four hours after the outburst began; Column 2: 0.3--10\,keV average absorbed flux (erg\,cm$^{-2}$\,s$^{-1}$) during a 50\,ks observation of a GRB afterglow with the X-IFU; Column 3: Minimum observed equivalent width (eV) of \ion{O}{vii} with a 90\% probability of detection at a minimum significance of 90\% by the X-IFU; Column 4: Minimum observed equivalent width (eV) of \ion{O}{vii} with a 50\% probability of detection at a minimum significance of 90\% by the X-IFU; Column 5: Minimum observed equivalent width (eV) of \ion{O}{vii} with a 20\% probability of detection at a minimum significance of 90\% by the X-IFU.}
\end{table}

\subsection{Recovery of equivalent width} \label{sec:equiv_widths}
The equivalent width that is fitted to the \ion{O}{vii} and \ion{O}{viii} absorption features in the simulated \textit{Athena} spectrum is of particular interest because it shows the capability of the X-IFU to reproduce and to measure the characteristics of narrow absorption features. 
The equivalent width is calculated from the \texttt{normalisation} parameter of the \texttt{zgauss} component of the fitted spectral model for all \ion{O}{vii}-\ion{O}{viii} absorption line pairs that were detected with at least a 90\% significance and were detected at the correct redshift during the analysis. 

\begin{figure}
\centering
\resizebox{\hsize}{!}{\includegraphics{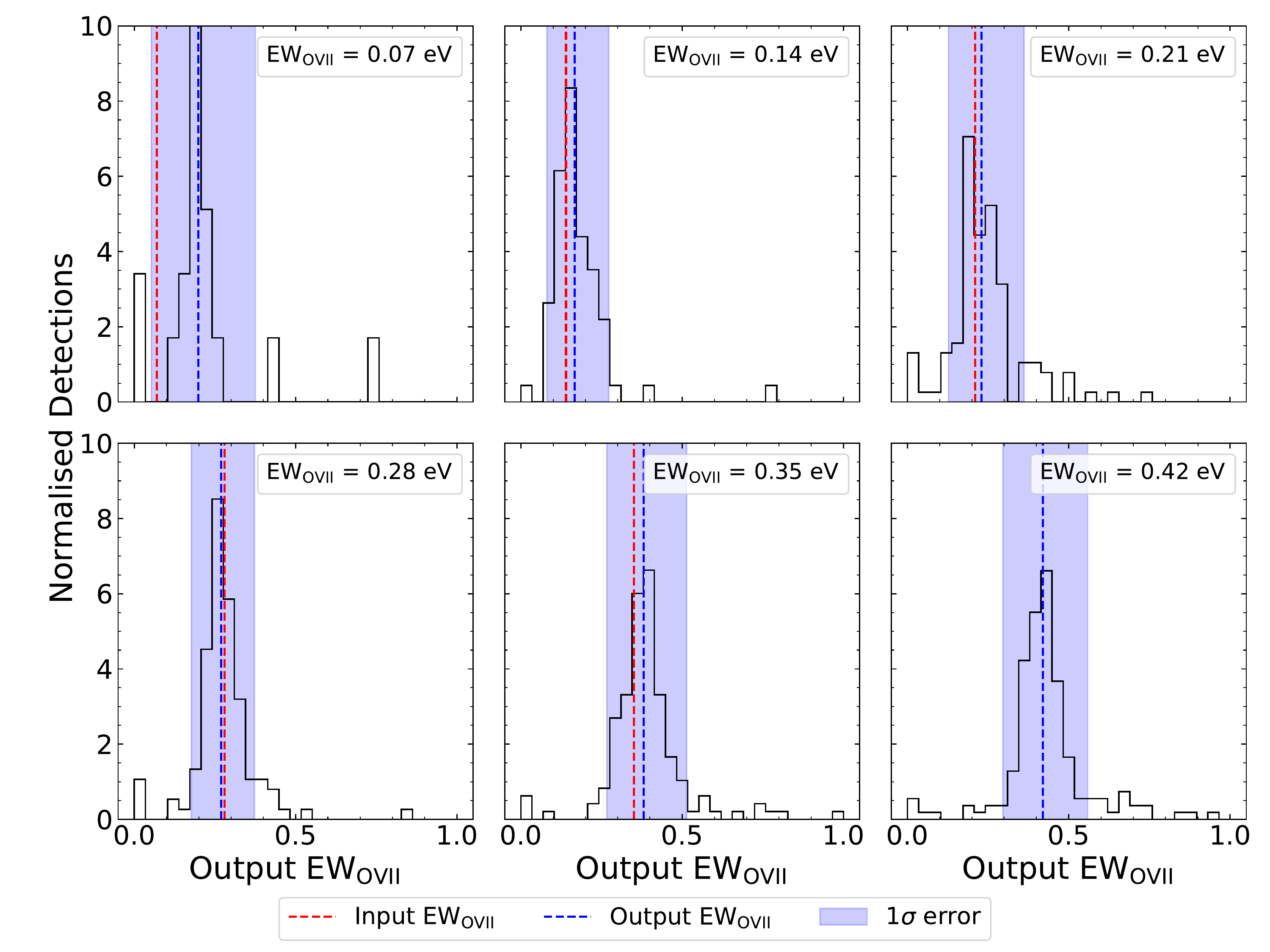}}
\caption{Distributions of the fitted \ion{O}{vii} equivalent widths in eV for each input \ion{O}{vii} equivalent width from GRB afterglow spectra with $F$\textsubscript{start}\,=\,10$^{-12}$--10$^{-10}$\,erg\,cm$^{-2}$\,s$^{-1}$. The input equivalent width is represented by the red dashed line and the median equivalent width of the distribution is shown by the blue dashed line. The blue shaded area is the 1$\sigma$ interval of the distribution.}
\label{fig:equiv_width_dist}
\end{figure}

\begin{figure}
\centering
\resizebox{\hsize}{!}{\includegraphics{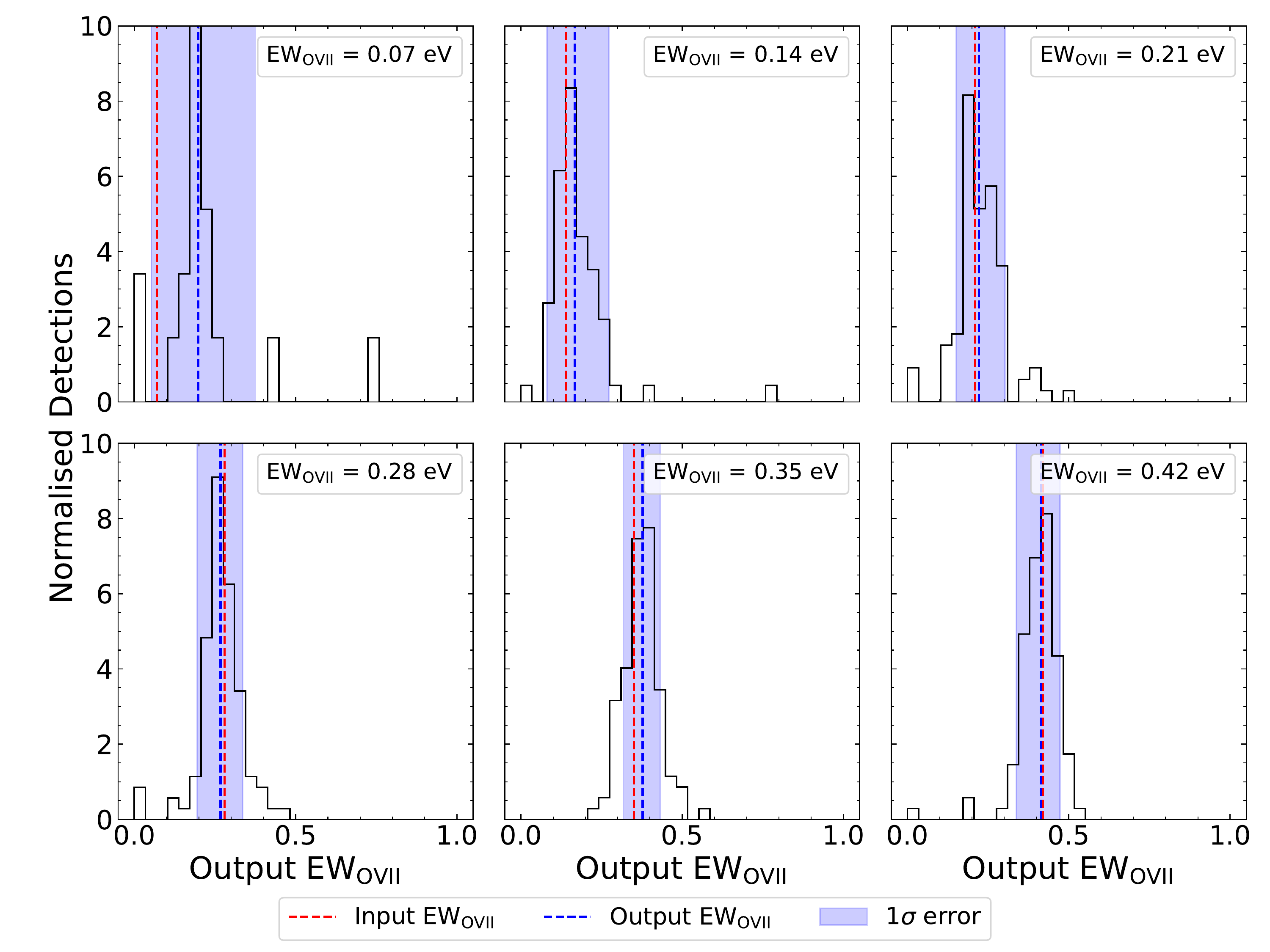}}
\caption{Distributions of the fitted \ion{O}{vii} equivalent widths in eV for each input \ion{O}{vii} equivalent width from GRB afterglow spectra with \textit{F}\,$\geq$\,5\,$\times$\,10$^{-11}$\,erg\,cm$^{-2}$\,s$^{-1}$. The input equivalent width is represented by the red dashed line and the median equivalent width of the distribution is shown by the blue dashed line. The blue shaded area is the 1$\sigma$ interval of the distribution.}
\label{fig:equiv_width_dist_bright}
\end{figure}

The \ion{O}{vii} equivalent width distributions from the fitted simulated spectra for all fluxes are shown in Fig.~\ref{fig:equiv_width_dist}. 
Each distribution consists of the \ion{O}{vii} equivalent width that was fitted during the blind line search for significantly detected absorption features of a given input equivalent width at the correct redshift, $z\textsubscript{WHIM}$. 
The median of the measured equivalent width distribution is marked by a blue-dash line while the input equivalent width is marked by a red-dash line.
The 1$\sigma$ error of the distribution is shaded in blue.
The figure shows that for EW\textsubscript{\ion{O}{vii}}\,$\geq$\,0.14\,eV, the X-IFU can recover the equivalent width of the input lines to within 1$\sigma$ of the observed equivalent width.
However, all distributions in Fig.~\ref{fig:equiv_width_dist} contain outliers to the 1$\sigma$ range of the input equivalent width.
These outliers arise from real absorption features detected in afterglow spectra with \textit{F}\textsubscript{start}\,<\,5\,$\times$\,10$^{-11}$\,erg\,cm$^{-2}$\,s$^{-1}$.
For line equivalent widths of EW\textsubscript{\ion{O}{vii}}\,$\geq$\,0.35\,eV, Fig.~\ref{fig:35mm_detect} shows that a fainter GRB afterglow of 10$^{-11}$\,erg\,cm$^{-2}$\,s$^{-1}$ will give at least a 70\% chance of detecting the absorber in comparison to less than a 20\% chance for weaker absorbers.
Therefore, the distributions in Fig.~\ref{fig:equiv_width_dist} with high equivalent widths contain measurements from significant detections of absorption features in fainter afterglow spectra. 
The low flux of the afterglow results in the equivalent widths of the absorption lines not being measured within the 1$\sigma$ range of the input equivalent width.
Figure~\ref{fig:equiv_width_dist_bright} shows these equivalent width distributions when detections made from spectra with \textit{F}\textsubscript{start}\,<\,5\,$\times$\,10$^{-11}$\,erg\,cm$^{-2}$\,s$^{-1}$ are excluded.
The distributions show fewer outliers outside the 1$\sigma$ range in comparison to those present in Fig.~\ref{fig:equiv_width_dist}, while the recovered equivalent width lies closer to the input.

The distributions for absorbers of low equivalent widths, in particular EW$_\mathrm{OVII}$\,=\,0.07\,eV, are visually poorer than those of high equivalent widths since there are less significant detections made at low equivalent widths. 
Figure~\ref{fig:35mm_detect} shows that the probability of detection for EW$_\mathrm{OVII}$\,=\,0.07\,eV remains low for all fluxes, reaching a maximum of approximately 30\% for the highest flux simulated. 
In comparison, EW$_\mathrm{OVII}$\,$\geq$\,0.21\,eV reach a near 100\% probability of detection for $F_{start}$\,$\geq$\,5$\times$10$^{-11}$\,erg\,cm$^{-2}$\,s$^{-1}$. 
The number of detections in each distribution of Fig.~\ref{fig:equiv_width_dist} is 17, 66, 111, 109, 140, and 159 for EW$_\mathrm{OVII}$\,=\,0.07, 0.14, 0.21, 0.28, 0.35, and 0.42\,eV, respectively.
Therefore, the sample size of each distribution varies due to the ability of the X-IFU to detect significant absorption features of high equivalent widths with a much higher probability. 

The distributions of Figs.~\ref{fig:equiv_width_dist} and \ref{fig:equiv_width_dist_bright} exhibit a positive shift between the median of the detected equivalent widths and the input equivalent width from the simulated spectra.
Each distribution exhibits broadening about the input equivalent width due to the Eddington bias \citep{1913MNRAS..73..359E,1940MNRAS.100..354E} and, therefore, contains simulated absorption features of a mean equivalent width, EW with a variance, $\Delta$EW. 
Since the distributions consist only of \ion{O}{vii} features that were detected above a 90\% significance level at the correct absorber redshift during the blind-line fitting process, the positive shift between the median and the input equivalent width of each distribution increases with decreasing equivalent width.
For higher equivalent widths that are well above the detector threshold, there is an equal probability that the equivalent width of a significant feature will be detected with EW\,+\,$\Delta$EW or EW\,--\,$\Delta$EW.
However, as the detector threshold is approached it becomes more likely that a significant detection is made for EW\,+\,$\Delta$EW, thus introducing a positive shift in the observed distributions for lines with low equivalent widths.
The EW\textsubscript{\ion{O}{vii}}\,=\,0.07\,eV distribution of both figures samples very few absorbers because of its low input equivalent width, resulting in few significant detections being made during the blind line analysis.
It can be concluded that the equivalent width of these absorption features is too small to be accurately measured with the GRB afterglow fluxes simulated in this paper. 

\begin{figure*}
\centering
\includegraphics[width=14cm]{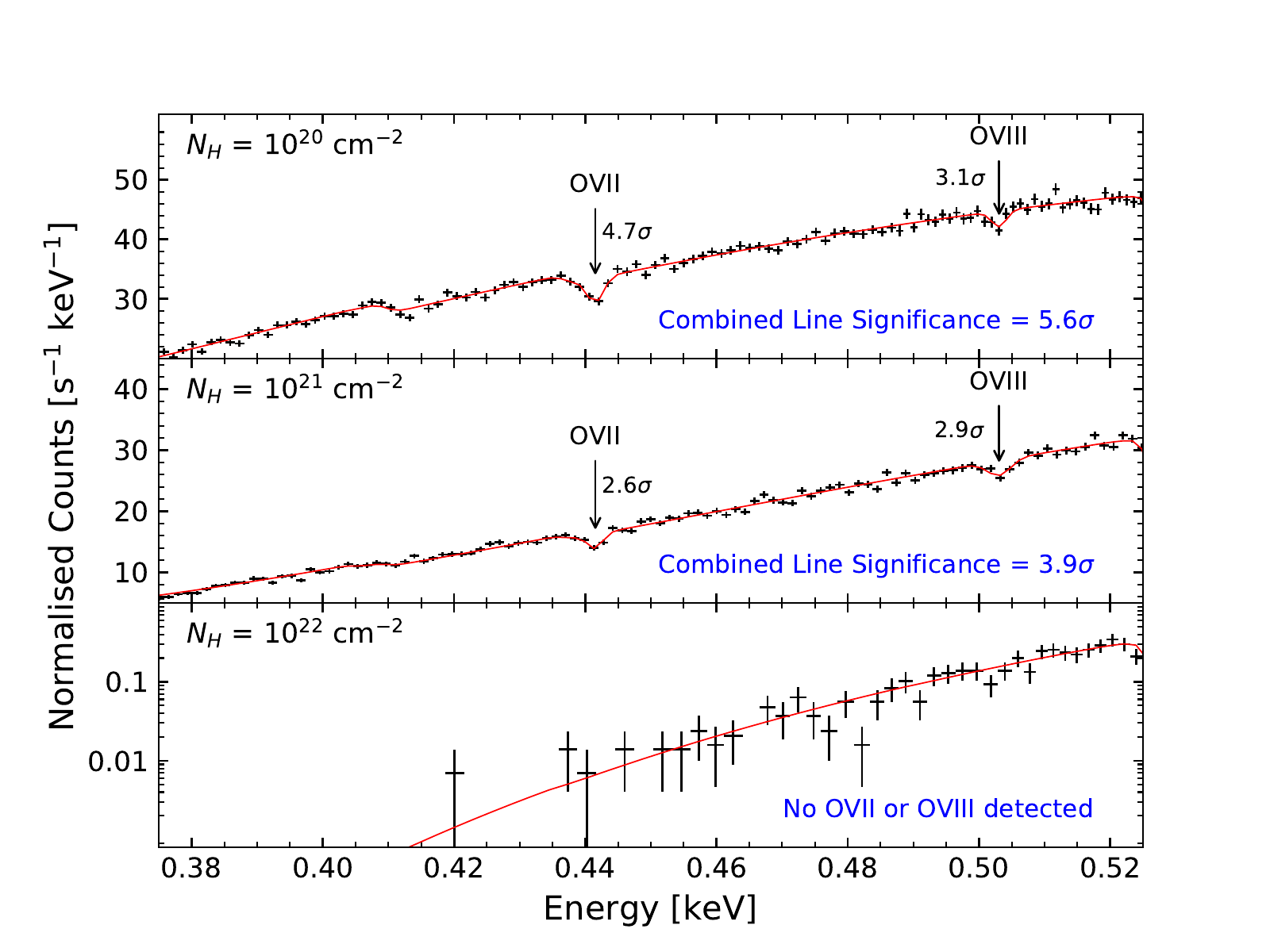}
\caption{Simulated X-IFU spectrum for varying Milky Way hydrogen column densities with \ion{O}{vii} and \ion{O}{viii} absorption features at $z$\textsubscript{WHIM}\,=\,0.3. Input lines have equivalent widths of EW\textsubscript{\ion{O}{vii}}\,=\,0.28\,eV and EW\textsubscript{\ion{O}{viii}}\,=\,0.19\,eV. Afterglow spectra have a starting flux of $F$\textsubscript{start}\,=\,5\,$\times$\,10$^{-11}$\,erg\,cm$^{-2}$\,s$^{-1}$. \textit{Top:} Hydrogen column density of 10$^{20}$\,cm$^{-2}$. \textit{Middle:} Hydrogen column density of 10$^{21}$\,cm$^{-2}$. \textit{Bottom:} Hydrogen column density of 10$^{22}$\,cm$^{-2}$. }
\label{fig:nh}
\end{figure*}

\subsection{Effect of galactic absorption on the detection of \ion{O}{vii} and \ion{O}{viii}} \label{sec:nh_effect}
The rest energies of \ion{O}{vii} and \ion{O}{viii} (574\,eV and 654\,eV, respectively) occur near the 200\,eV low energy threshold of the X-IFU instrument. 
The energies of their absorption features are redshifted to lower energies by the redshift of the originating absorber. 
As GRBs occur at an average redshift of 2, it is possible to obtain absorption features from the WHIM in afterglow spectra up to that redshift. 
However, once a WHIM filament reaches a redshift of $z$\textsubscript{WHIM}\,=\,1, an \ion{O}{vii} absorption feature will occur at an energy of 287\,eV while \ion{O}{viii} will be at 327\,eV. 
While these energies are still within the energy range of the X-IFU, they occur at energies within the afterglow spectrum that are significantly absorbed by the interstellar medium within the Milky Way.

To assess the impact that Galactic absorption along the line of sight has on the detection of \ion{O}{vii} and \ion{O}{viii}, SIXTE simulations of a GRB afterglow were executed for varying hydrogen column densities within the Milky Way galaxy and for varying WHIM redshifts, $z$\textsubscript{WHIM}. 
The properties of the GRB afterglow are as in Table \ref{tab:model_components}, while the Galactic absorption parameter of the local galaxy is varied from 10$^{20}-$10$^{22}$\,cm$^{-2}$.
This range represents the varied distribution of $N_\mathrm{H}$ columns in the Milky Way, where typical values of 10$^{22}$\,cm$^{-2}$ originate within the Galactic plane and lower values of 10$^{20}$\,cm$^{-2}$ occur at regions of high Galactic latitude.
The redshift of the absorption features is varied from $z$\textsubscript{WHIM}\,=\,0--0.5.
The \ion{O}{vii} absorption lines have an equivalent width of 0.28\,eV while the equivalent width of \ion{O}{viii} is 0.19\,eV. 
The starting flux of the afterglow is set to 5\,$\times$\,10$^{-11}$\,erg\,cm$^{-2}$\,s$^{-1}$. 
From the results presented in Sect.~\ref{sec:35mm}, this combination of equivalent width and starting flux should provide a detection probability of both lines close to unity, based on a Galactic hydrogen column density of 2\,$\times$\,10$^{20}$\,cm$^{-2}$. 
The spectra are analysed using the blind line fitting technique of Sect.~\ref{sec:spectral_fitting} and if the line detection is determined significant at at least a 90\% threshold and is located at the correct redshift, the true single line significance of each line is calculated using the normalisation parameter, \textit{K} of the Gaussian absorption component by \textit{K}/$\Delta$\textit{K} \citep{2018arXiv181002207B}.
The combined line significance of the \ion{O}{vii}-\ion{O}{viii} pair is calculated by adding each single line significance in quadrature. 
For each combination of hydrogen column density and WHIM redshift, ten spectra are simulated.
The combined line significance from all ten spectra are used to determine the average significance and the 1$\sigma$ error for each column density and redshift.

Three simulated spectra are shown in Fig.~\ref{fig:nh}. The spectra show a small energy range around the \ion{O}{vii} and \ion{O}{viii} absorption features, originating from a WHIM absorber at $z$\textsubscript{WHIM}\,=\,0.3.
For $N_\mathrm{H}$\,=\,10$^{20}$\,cm$^{-2}$, the single line significance of \ion{O}{vii} is calculated to be 4.7$\sigma$ while \ion{O}{viii} has a calculated single line significance of 3.1$\sigma$, giving a combined statistical significance of 5.6$\sigma$. 
The \ion{O}{vii} absorption feature is detected at slightly higher significance than \ion{O}{viii} because of its higher input equivalent width. 
For $N_\mathrm{H}$\,=\,10$^{21}$\,cm$^{-2}$, the single line significance of \ion{O}{vii} and \ion{O}{viii} are 2.6$\sigma$ and 2.9$\sigma$, respectively, giving a combined significance of 3.9$\sigma$.
The detection significance of the lines has decreased with the increase in the hydrogen column density simulated owing to the excess absorption occurring at the energy of the lines. 
Inspite of the higher input equivalent width into the simulated spectra of \ion{O}{vii} over \ion{O}{viii}, the higher absorption from the local galaxy causes it to be detected at a slightly lower significance than \ion{O}{viii}.
For the highest simulated hydrogen column density of $N_\mathrm{H}$\,=\,10$^{22}$\,cm$^{-2}$, the X-IFU instrument does not detect any counts below $\approx$0.42\,keV. 
Neither \ion{O}{vii} nor \ion{O}{viii} are detected significantly within the spectrum because of the high level of Galactic absorption.
These spectra have in excess of 10$^{6}$ high resolution X-IFU counts in the 0.3--10\,keV energy range, suggesting that a detection should be made following the results presented in Fig.~\ref{fig:35mm_detect}. 
However, a detection will not be deemed significant in spectra with a high hydrogen column density if the total number of high resolution X-IFU counts in the energy range of the \ion{O}{vii}-\ion{O}{viii} pair is too low, as in the bottom panel of Fig.~\ref{fig:nh}.
Therefore, a bright afterglow spectrum with a high count rate over the full X-IFU energy range may not detect absorption features at the soft X-ray energy range of the spectrum due to extreme local hydrogen absorption along the line of sight to the GRB.

\begin{figure}
\centering
\resizebox{\hsize}{!}{\includegraphics{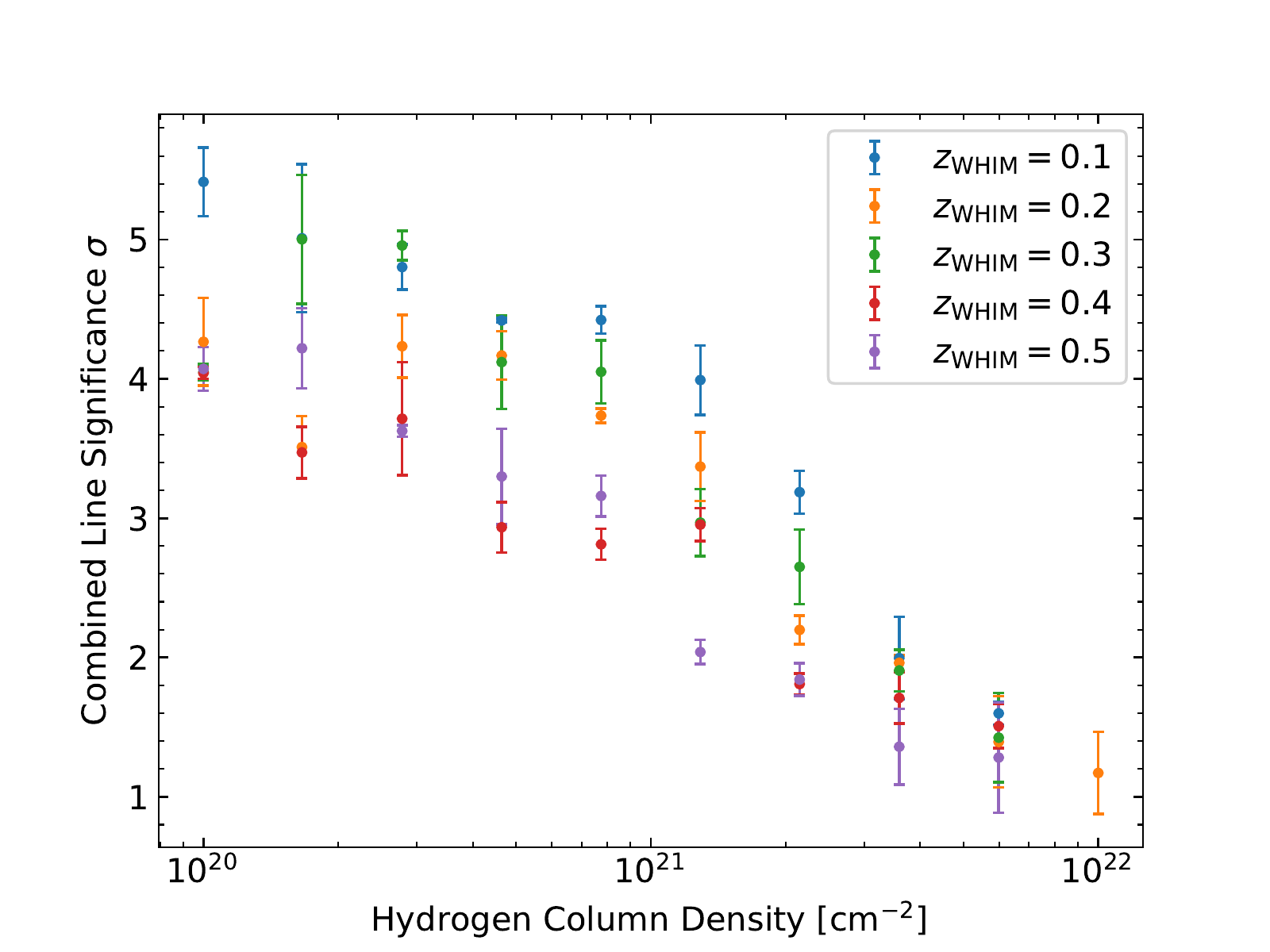}}
\caption{Combined line significance of an \ion{O}{vii}-\ion{O}{viii} pair detection for different WHIM absorber redshifts as a function of the hydrogen column density in the Milky Way galaxy for an afterglow flux of \textit{F}\textsubscript{start}\,=\,5\,$\times$\,10$^{-11}$\,erg\,cm$^{-2}$\,s$^{-1}$and \ion{O}{vii} line equivalent width of EW\textsubscript{\ion{O}{vii}}\,=\,0.28\,eV. Error bars represent the 1$\sigma$ scatter among the 10 simulated spectra.}
\label{fig:sigma_vs_nh}
\end{figure}

Figure \ref{fig:sigma_vs_nh} shows the effect of the Galactic hydrogen column density on the detection significance of an \ion{O}{vii}-\ion{O}{viii} absorption pair.
As the column density of hydrogen in the local galaxy increases, the combined line significance of \ion{O}{vii}-\ion{O}{viii} decreases. 
Increasing the redshift of the WHIM absorption also results in a reduction in the detection significance.
For higher redshifts, the lines occur in the afterglow spectrum at softer energies where the absorption from local hydrogen is more prominent, resulting in a lower line significance.
A high local hydrogen column density and a high redshift can result in \ion{O}{vii}, or both \ion{O}{vii} and \ion{O}{viii}, not being detectable in the spectrum, as seen in the bottom panel of Fig.~\ref{fig:nh}.
However, \ion{O}{viii} may still be detected.
In this case, the combined line significance is simply the true significance of the \ion{O}{viii} absorption feature.

The number of independent resolution elements $N$ in the blind line search of the spectrum is calculated as in \citet{2018arXiv181002207B} by $N = E_{\ion{O}{vii}}\,\cdot\,\Delta\,z\textsubscript{WHIM} / \sigma_{E}$, where E\textsubscript{{\ion{O}{vii}}} is the rest energy of \ion{O}{vii}, $\Delta$\,z\textsubscript{WHIM} is the redshift range in which absorption features are searched for, and $\sigma_{E}$ is the energy spectral resolution of the X-IFU.
The X-IFU spectral resolution of $\sigma_{E}$\,=\,2.5\,eV and the WHIM redshift range of $\Delta z\textsubscript{WHIM}$\,=\,0.5 gives 115 independent opportunities to detect a feature during the search.
For each absorption feature to have a true line significance of at least 3$\sigma$, a minimum combined line significance of 4.2$\sigma$ is required. 
Figure \ref{fig:sigma_vs_nh} shows that to obtain a detection with a combined line significance of 4.2$\sigma$ or greater, the hydrogen column density of the local galaxy must be less than $\sim$10$^{21}$\,cm$^{-2}$.
From \citet{2018arXiv181002207B} the probability of detecting two random fluctuations both of 3$\sigma$ in a blind line search having 100 independent trials  is 3.67\%, meaning it could be concluded that the detected fluctuations are random.
However, if both fluctuations are detected at 3.2$\sigma$ in a blind line search having the same number of trials, the probability that the fluctuations are random is reduced to 0.4\%, meaning there is a much greater chance that these fluctuations are real.
Two 3.2$\sigma$ detections result in a combined line significance of 4.5$\sigma$.
At this significance, Fig.~\ref{fig:sigma_vs_nh} shows that the hydrogen column density must be less than 8\,$\times$\,10$^{20}$\,cm$^{-2}$.

\section{Discussion} \label{sec:discussion}
\begin{figure*}
\centering
\includegraphics[width=17cm]{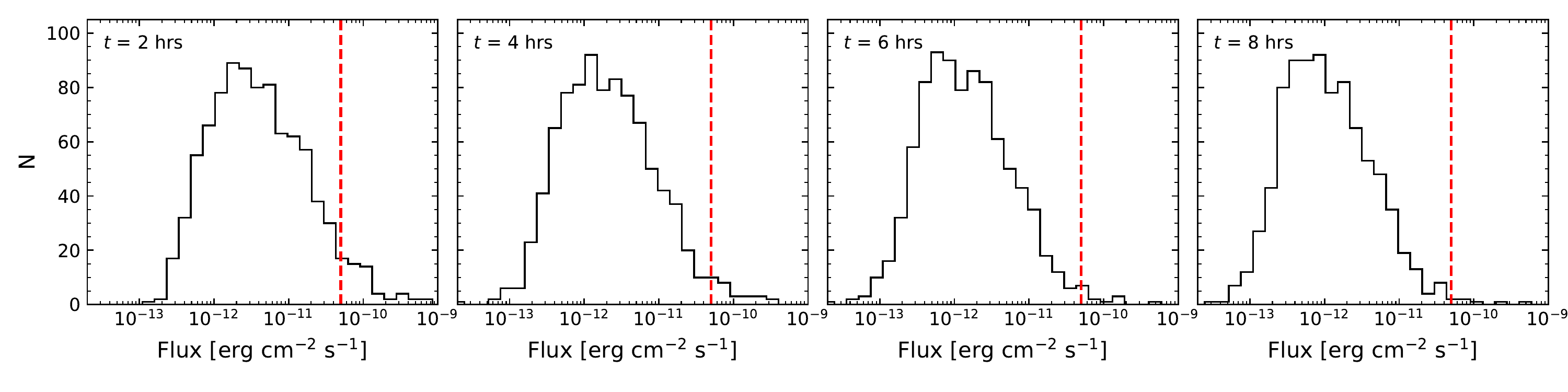}
\caption{Flux distribution of GRBs observed by \textit{Swift}-XRT with T$_{90}$\,>\,4\,s (long bursts only) at two hours (left), four hours (middle left), six hours (middle right), and eight hours (right) after the initial outburst began. The red dashed line indicates a flux of 5\,$\times$\,10$^{-11}$\,erg\,cm$^{-2}$\,s$^{-1}$.}
\label{fig:flux_distributions}
\end{figure*}

If GRBs are used to detect the WHIM with a minimum limit of 75\% confidence of detection, \textit{Athena} will need to target GRB afterglows of \textit{F}\textsubscript{start}\,>\,5\,$\times$\,10$^{-11}$\,erg\,cm$^{-2}$\,s$^{-1}$ to detect an \ion{O}{vii}-\ion{O}{viii} pair of minimum equivalent width 0.18\,eV and 0.12\,eV, respectively. 
This flux provides a false-alarm probability of less than 10\% for detecting the WHIM with lines of this equivalent width.
Strong \ion{O}{vii} absorbers of 0.39--0.42\,eV can be detected with a probability of 75\% at 10$^{-11}$\,erg\,cm$^{-2}$\,s$^{-1}$ and a false-alarm probability of 20\%.
Brighter afterglows of \textit{F}\textsubscript{start}\,$\sim$\,10$^{-10}$\,erg\,cm$^{-2}$\,s$^{-1}$ will be required to detect \ion{O}{vii} and \ion{O}{viii} absorption features of 0.13\,eV and of 0.08\,eV equivalent width, respectively. 
These starting flux thresholds, and corresponding X-IFU count rates and 0.3--10\,keV average absorbed fluxes, are listed in the first, second, and third Columns of Table \ref{tab:number_absorbers}, while the minimum detectable equivalent widths of the \ion{O}{vii}-\ion{O}{viii} pair are listed in Columns 5 and 7. 

To determine how many GRB afterglows occur per year in line with the flux thresholds of Table \ref{tab:number_absorbers} and, therefore, allow the detection of the WHIM with \textit{Athena}, the \textit{Swift}-XRT population of GRB afterglows is considered \citep{2007A&A...469..379E, 2009MNRAS.397.1177E}.
From the \textit{Swift} population all long GRBs are considered, with a limit of T$_{90}$\,>\,4\,s to exclude any short GRBs with extended emission, producing a sample of 948 GRB afterglows.
The light curves of the afterglows are interpolated and are assessed to determine a flux value at a given observation time. If no interpolated flux value is present for the observation time, it is excluded from the sample.
Figure \ref{fig:flux_distributions} shows the flux distributions of 903, 892, 889, and 889 GRB afterglows at two, four, six, and eight hours, respectively, after the initial outburst began.

The original ToO requirement for \textit{Athena} aimed to observe 67\% of GRB afterglows in a 60\% field of regard within four hours for 50\,ks.
However, to account for the decaying nature of GRB afterglow fluence with time the requirement has been reformulated.
The science requirement now aims to observe GRB afterglows that can provide spectra with at least one million counts \citep{2019IWPSS_Jaubert}.
The reformulation allows the X-IFU to target dimmer afterglows early in their afterglow phase or target brighter afterglows at later times, provided that the minimum number of spectral counts are obtained.
This can be achieved by optimizing the cooling chain scheme of the X-IFU with a partial heat-up (PHU) operational strategy, whereby the regeneration of the cooling chain between X-IFU observations is split into two sub-phases.
The first heat-up (HU) phase begins immediately after an X-IFU observation and occurs for a duration that provides the detector with sufficient cold time to perform a potential ToO observation that would meet the required count rate.
Therefore, depending on the magnitude of repointing required, the brightness of the afterglow, and the operational phase of the X-IFU cooling chain, the reaction time of \textit{Athena} will vary.
For an \textit{Athena} ToO reaction time of four hours, the four hour distribution of Fig.~\ref{fig:flux_distributions} shows that the population of GRBs higher than the threshold starting fluxes of 1, 5, and 10 $\times$\,10$^{-11}$\,erg\,cm$^{-2}$\,s$^{-1}$ are 15\%, 3\%, and 1\% of the total.  
These fluxes allow for the detection of absorption features with EW\textsubscript{\ion{O}{vii}}\,$\geq$\,0.39, 0.18, and 0.13\,eV, respectively. 
To detect lines of EW\textsubscript{\ion{O}{vii}}\,<\,0.13\,eV, a brighter afterglow would be required.
However, less than 1\% of the total population of GRB afterglows detected by \textit{Swift} have a flux of $F$\,\,>\,10$^{-10}$\,erg\,cm$^{-2}$\,s$^{-1}$ four hours after the initial outburst, and so, it is unlikely that such a burst would contribute to the detection of the WHIM. 
Alternatively, a brighter afterglow could be seen if the X-IFU observation begins sooner than four hours after trigger. 
If \textit{Athena} can react to a GRB trigger and repoint at the afterglow within two hours, the population of afterglows with starting fluxes greater than 10$^{-10}$\,erg\,cm$^{-2}$\,s$^{-1}$ increases to 3\%.
This allows for a more likely detection of weaker absorption systems with EW\textsubscript{\ion{O}{vii}}\,<\,0.13\,eV.
A longer reaction time of eight hours would result in the GRB afterglow population decreasing to 6\%, 1\%, and 0.5\% for starting fluxes of 1, 5, and 10 $\times$\,10$^{-11}$\,erg\,cm$^{-2}$\,s$^{-1}$, meaning that a significant detection of the WHIM becomes unlikely, but not impossible. 

The exact number of GRBs that \textit{Athena} will detect per year will depend on how the bursts are triggered. 
Currently, $\gamma$-ray observatories such as the Neils Gehrels Swift Observatory \citep{2004ApJ...611.1005G}, \textit{Fermi}-GBM \citep{2009ApJ...702..791M}, and INTEGRAL \citep{2003A&A...411L...1W} trigger on and perform follow-up observations of GRBs.
They provide redshift and localisation information to X-ray observatories to facilitate further X-ray afterglow follow up.
However, these missions have all exceeded their nominal mission lifetime.
Future missions with triggering capabilities planned to be operating during the lifetime of \textit{Athena} include the Space Variable Objects Monitor \citep[SVOM;][]{2015arXiv151203323C} and the Wide Field Monitor \citep{2018SPIE10699E..48H} on board the Enhanced X-ray Timing and Polarization mission \citep[eXTP;][]{2016SPIE.9905E..1QZ}.
A number of proposed missions are currently at concept level, such as the Transient High Energy Sky and Early Universe Surveyor \citep[THESEUS;][]{2018AdSpR..62..191A} and the All-Sky Medium Energy Gamma-ray Observatory \citep[AMEGO;][]{2019arXiv190707558M}, both capable of triggering and localising GRBs. 
In addition, constellations of low-cost CubeSat missions capable of detecting gamma-rays, such as BurstCube \citep{2017arXiv170809292R} and HERMES \citep{2019NIMPA.936..199F}, could provide trigger and localisation information.
A mission with such capabilities is required to increase the rate of GRB detections per year with accurate localisations and redshifts, so that rapid repointing of the \textit{Athena} X-IFU can be achieved to conduct high resolution X-ray spectroscopy of the WHIM.

\begin{figure}
\centering
\resizebox{\hsize}{!}{\includegraphics{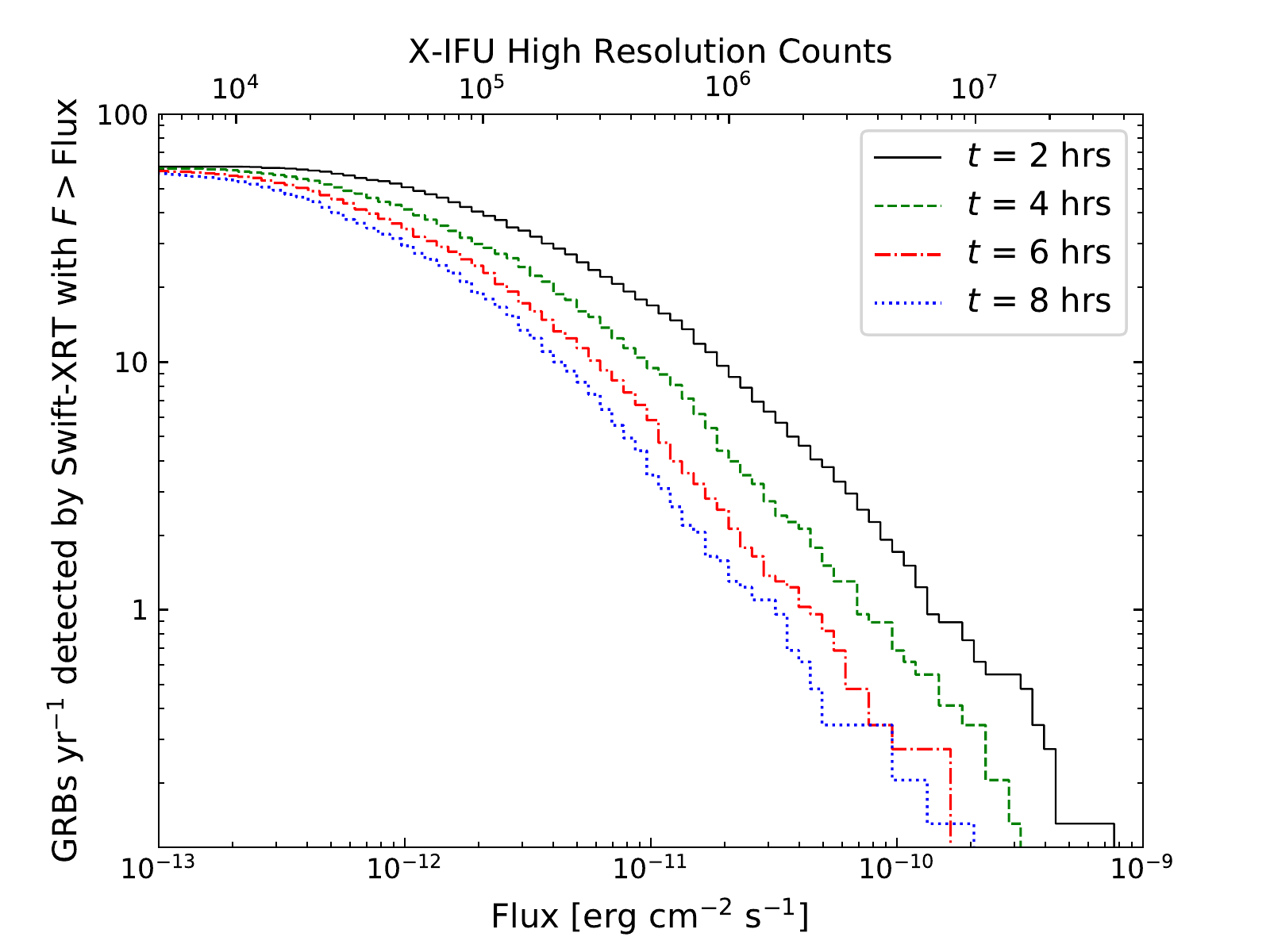}}
\caption{Cumulative number of GRBs afterglows detected per year by \textit{Swift}-XRT with T$_{90}$\,>\,4\,s (long bursts only) as a function of flux and X-IFU counts for observations made at two, four, six, and eight hours after the initial outburst began.}
\label{fig:cumulative_distributions}
\end{figure}

Table~\ref{tab:number_absorbers} presents the detection parameters for \ion{O}{vii}-\ion{O}{viii} absorption features against GRB afterglows of varying starting fluxes. 
Column 5 of Table \ref{tab:number_absorbers} reports the minimum observed equivalent width required to detect an \ion{O}{vii} absorption feature of an \ion{O}{vii}-\ion{O}{viii} pair for each afterglow flux in Column 1, with the \ion{O}{viii} minimum equivalent width of that same pair in Column 9. 
Columns 6, 7, and 8 report the number of \ion{O}{vii} absorption systems per unit redshift above the given X-IFU observed equivalent width threshold for distributions of the WHIM at $z$\,=\,0.1, 0.2, and 0.5, respectively.
Similarly, Columns 10, 11, and 12 report the number of \ion{O}{viii} absorption systems per unit redshift above the observed equivalent width threshold in Column 9.
These are calculated from the EAGLE cosmological hydrodynamical simulations \citep{2015MNRAS.446..521S, 2015MNRAS.450.1937C, 2016A&C....15...72M}, using the methods described in \citet{2019MNRAS.488.2947W}.
The number of \ion{O}{vii} and \ion{O}{viii} absorbers decreases with increasing redshift.
This is a result of the rest frame equivalent width increasing with increasing redshift, for a fixed observed equivalent width measured in energy. 
The number of absorbers at $z$\,=\,0.1 and at $z$\,=\,0.2 show little evolution, while the decrease in absorbers becomes prominent at $z$\,=\,0.5.

\begin{table*}
\centering
\caption{Detection parameters of the WHIM for the X-IFU.}
\begin{tabular}{c c c c c c c c c c c c}
\hline\hline
\multicolumn{1}{c}{$F$\textsubscript{start}}
& \multicolumn{1}{c}{\centering $F_{0.3-10}$}
& \multicolumn{1}{c}{\centering \small X-IFU Counts}
& \multicolumn{1}{c}{\# \small{GRBs}}
& \multicolumn{1}{c}{\centering \small EW\textsubscript{\ion{O}{vii}}}
& \multicolumn{3}{c}{d$N$\,(\small {$>$\,EW\textsubscript{{\ion{O}{vii}}}})\,/\,d$z$}
& \multicolumn{1}{c}{\centering \small EW\textsubscript{\ion{O}{viii}}}
& \multicolumn{3}{c}{d$N$\,(\small {$>$\,EW\textsubscript{{\ion{O}{viii}}}})\,/\,d$z$}
\\ \multicolumn{1}{c}{{\footnotesize[erg\,cm$^{-2}$\,s$^{-1}$]}}
& \multicolumn{1}{c}{{\footnotesize[erg\,cm$^{-2}$\,s$^{-1}$]}}
& \multicolumn{1}{c}{\centering {\footnotesize(in 50\,ks)}}
& \multicolumn{1}{c}{\%}
& \multicolumn{1}{c}{{\footnotesize[eV]}}
& \multicolumn{1}{c}{\footnotesize{$z=0.1$}} & \multicolumn{1}{c}{\footnotesize{$z=0.2$}} & \multicolumn{1}{c}{\footnotesize{$z=0.5$}}
& \multicolumn{1}{c}{{\footnotesize[eV]}}
&\multicolumn{1}{c}{\footnotesize{$z=0.1$}} & \multicolumn{1}{c}{\footnotesize{$z=0.2$}} & \multicolumn{1}{c}{\footnotesize{$z=0.5$}}
\\ \hline
1\,$\times$\,10$^{-10}$  & 3.3\,$\times$\,10$^{-11}$ & 4.8\,$\times$\,10$^{6}$  & 1.46  & 0.13 & 2.39 & 2.40 & 1.81 & 0.09 & 2.32    & 2.41    & 2.44 \\
5\,$\times$\,10$^{-11}$  & 1.7\,$\times$\,10$^{-11}$ & 2.4\,$\times$\,10$^{6}$  & 2.8  & 0.18 & 1.33 & 1.32 & 0.85 & 0.12 & 1.46    & 1.50    & 1.46 \\
4\,$\times$\,10$^{-11}$  & 1.4\,$\times$\,10$^{-11}$ & 1.9\,$\times$\,10$^{6}$  & 3.81  & 0.20 & 1.04 & 1.03 & 0.63 & 0.13 & 1.25    & 1.29    & 1.24 \\
3\,$\times$\,10$^{-11}$  & 1.0\,$\times$\,10$^{-11}$ & 1.4\,$\times$\,10$^{6}$  & 4.59  & 0.30 & 0.30 & 0.31 & 0.14 & 0.20 & 0.48    & 0.47    & 0.37  \\
1\,$\times$\,10$^{-11}$  & 3.3\,$\times$\,10$^{-12}$ & 4.8\,$\times$\,10$^{5}$  & 15.58  & 0.39 & 0.11 & 0.13 & 0.04 & 0.26 & 0.20    & 0.18    & 0.12 \\
\hline
\end{tabular}
\label{tab:number_absorbers}
\tablefoot{Column 1: simulated starting flux (erg\,cm$^{-2}$\,s$^{-1}$) of the GRB afterglow; Column 2: 0.3--10\,keV average absorbed flux (erg\,cm$^{-2}$\,s$^{-1}$) during a 50\,ks observation of a GRB afterglow with the X-IFU; Column 3: High resolution counts collected during the 50\,ks observation of the GRB afterglow; Column 4: the percentage of GRB afterglows per year above the flux in Column 2 based on the Swift population of GRBs; Column 5: Minimum equivalent width (eV) of \ion{O}{vii} for a 75\% chance of detection with the X-IFU;
Column 6--8: Expected number of absorption systems per unit redshift for \ion{O}{vii} above EW in Column 5 at redshifts $z=$\,0.1, 0.2 and 0.5; Column 9: Minimum equivalent width (eV) of \ion{O}{viii} for a 75\% chance of detection with the X-IFU;
Column 10--12: Expected number of absorption systems per unit redshift for \ion{O}{viii} above EW in Column 9 at redshifts $z=$\,0.1, 0.2 and 0.5.}
\end{table*}

\begin{figure}
\centering
\resizebox{\hsize}{!}{\includegraphics{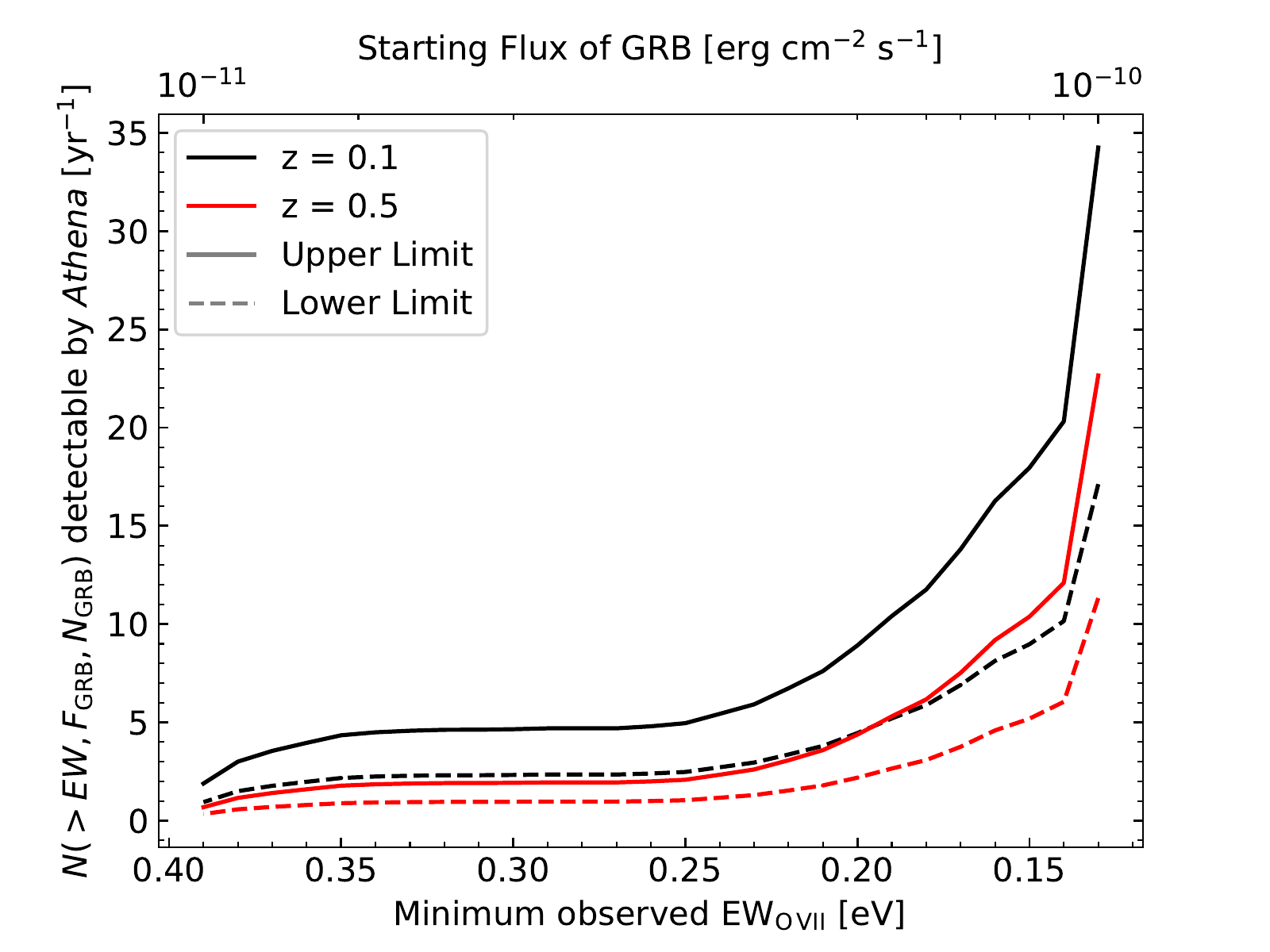}}
\caption{Cumulative number of \ion{O}{vii}-\ion{O}{viii} absorbers detectable by \textit{Athena} per year above a minimum observed equivalent width for distributions of the WHIM at $z=0.1$ (black) and $z=0.5$ (red) as a function of the starting flux of the GRB afterglow and the predicted number of afterglows above that flux.}
\label{fig:EW_distribution}
\end{figure}

Figure~\ref{fig:EW_distribution} shows the cumulative annual absorber detection rate of \textit{Athena}, based on the predicted GRB population and the reformulated ToO requirement.
The distribution is described by the function,
\begin{equation}
    N = \dfrac{dN(>EW)}{dz} \cdot dz \cdot N_\mathrm{Athena\,GRBs/yr} \cdot FoR \cdot N_\mathrm{H},
\end{equation}
where $dN(>EW)/dz$ is the number of absorbers per unit redshift above a minimum observed equivalent width given in Table~\ref{tab:number_absorbers}, $dz$ is the absorption length for the WHIM, $N_\mathrm{Athena\,GRBs/yr}$ is the predicted number of GRB afterglows that will be observed by \textit{Athena} per year to detect the WHIM, $FoR$\,=\,0.6 is the field of regard of \textit{Athena}, and $N_\mathrm{H}$ is the fraction of the field of regard with a suitable Galactic hydrogen column to observe the absorption features.
Given that the average redshift of a GRB is $z$\textsubscript{GRB}\,$\approx$\,2, it is assumed that the absorption length for the WHIM in a GRB observation will be $\Delta z\textsubscript{WHIM}$\,=\,1, as higher redshift absorption features are outside the X-IFU energy range.

Two distributions of the detected absorber rate are shown in Fig.~\ref{fig:EW_distribution}, highlighting the difference between the detection rates when probing different redshifts of the WHIM. 
At EW$\approx$0.35--0.39\,eV, the rate of GRB detection for afterglows with $F$\,>\,10$^{-11}$\,erg\,cm$^{-2}$\,s$^{-1}$ is approximately 15\% of the total population for a reaction time of four hours and approximately 30\% for a reaction time of two hours.
This results in the distribution of detected absorbers per year increasing slightly, even with a significant decrease in the number of absorbers per unit redshift.
Between EW\,$\approx$\,0.25--0.35\,eV the cumulative number of detected absorbers levels off, although there is an increasing number of absorbers with smaller equivalent widths, as the rate of GRB afterglows decreases with increasing flux.
The distributions show a significant increase in the number of absorbers detected per year between observed equivalent widths of 0.13--0.25\,eV.
This is a result of the high number of absorbers expected to have minimum equivalent widths in this range, in spite of the low population of high flux GRBs.

The predicted annual rate of GRB detection with THESEUS is 300--700\,GRBs\,yr$^{-1}$ \citep{2018AdSpR..62..191A}, a relative increase by a factor of $\approx$\,5--10 on the annual rate currently detected by \textit{Swift}.
Figure~\ref{fig:cumulative_distributions} shows the \textit{Swift}-XRT cumulative distribution of GRB afterglows as a function of the observation starting flux for observations beginning two, four, six, and eight hours after the outburst was triggered.
Assuming that \textit{Athena} can follow up on all GRBs within its field of regard triggered by a spacecraft such as THESEUS, the \textit{Swift}-XRT sample of afterglow observations can be used to provide an estimate for the number of afterglows that \textit{Athena} will observe.
This is calculated by  $N_\mathrm{Athena\,GRBs/yr}$\,=\,$N_\mathrm{THESEUS\,GRBs/yr} \times N_\mathrm{Swift\,GRBs/yr}$, where $N_\mathrm{Swift\,GRBs/yr}$ is derived from Fig.~\ref{fig:cumulative_distributions} by weighting each reaction time distribution in accordance with simulated \textit{Athena} ToO response times \citep{2019IWPSS_Jaubert} and $N_\mathrm{THESEUS\,GRBs/yr}$ is the relative number of GRB detections of THESEUS compared to \textit{Swift}.
This provides an upper and lower limit to the number of GRBs, and therefore, absorbers per year that \textit{Athena} may observe.
Figure~\ref{fig:EW_distribution} shows that the lower predicted GRB detection rate amounts to approximately 17 absorbers per year, while the higher predicted rate gives 34 absorbers per year detected with \textit{Athena} through \ion{O}{vii}-\ion{O}{viii} absorption pairs.
If $z$\,=\,0.5 contributes most to the survey path, the number of absorbers detected decreases to 11--22 per year.
This is a relative increase of 20--25\% on the absorbers detected per year in comparison to the previous ToO requirement to observe GRB afterglows for 50\,ks within four hours of the initial trigger. 

\citet{2019MNRAS.488.2947W} presented the relationship between the equivalent width and column density of \ion{O}{vii} and \ion{O}{viii} absorption systems.
A linear relation is present between the absorber equivalent width and column density for all but the highest column densities and equivalent widths, where the relationship becomes non-linear.
\citet{2019MNRAS.488.2947W} show that a high column density traces high-density gas, typically associated with the circumgalactic medium (CGM).
The relationship between the column density and the equivalent width of absorbers is highly scattered and the distinction of these parameters between the CGM and IGM is not well defined. 
Therefore, the number of absorbers quoted contain absorbers of the CGM together with filaments of the WHIM.
In addition, the work presented here simulates only \ion{O}{vii}-\ion{O}{viii} absorption pairs of a fixed ratio.
Spectra originating from WHIM absorbers may contain features from other ions, depending on the temperature and density of the gas, such as \ion{Ne}{IX}, \ion{Ne}{X}, \ion{Fe}{XVII}, \ion{Fe}{XVIII}, etc.
Searching for these ions in afterglow spectra will aid the detection probability of absorbers, and so, the number of absorbers detectable by \textit{Athena} quoted in this paper is of a lower limit.

The results discussed above refer to the number of absorption systems of a given equivalent width from afterglow spectra simulated with a local hydrogen column density of 2\,$\times$\,10$^{20}$\,cm$^{-2}$ and a hydrogen column density intrinsic to the GRB of 10$^{22}$\,cm$^{-2}$. 
\citet{2013MNRAS.431..394W} show that the $N_\mathrm{H}$ density distribution of GRB afterglows observed by \textit{Swift}-XRT peaks at 2\,$\times$\,10$^{21}$\,cm$^{-2}$, a factor of ten greater than the spectra simulated in this work to calculate the detection probability of the WHIM. 
Section \ref{sec:nh_effect} presented the impact of a higher hydrogen column density within the Milky Way on the detection of absorption features imprinted on an afterglow spectrum.
Figure \ref{fig:sigma_vs_nh} showed that to detect an \ion{O}{vii}-\ion{O}{viii} absorption pair originating from the same WHIM filament with a combined line significance greater than 4.5$\sigma$, the $N_\mathrm{H}$ column cannot exceed 8\,$\times$\,10$^{20}$\,cm$^{-2}$.
Various surveys have been performed to determine the hydrogen distribution in the Milky Way \citep[e.g.][]{1990ARA&A..28..215D,2005A&A...440..775K, 2016A&A...594A.116H}.
\citet{2016A&A...594A.116H} show that approximately 35$\%$ of lines of sight have column densities of $N_\mathrm{H}$\,>\,8\,$\times$\,10$^{20}$\,cm$^{-2}$ and are located between an average of 22$^{\circ}$ North and 24$^{\circ}$ South of the galactic plane.
Because of the effect that such levels of absorption have on the detection of WHIM features, these regions of the sky should be excluded when searching for WHIM absorption features in GRB afterglow spectra.
This results in a 25\% reduction of the number of absorption systems that can be detected by \textit{Athena} within the field of regard, which is included in Fig~\ref{fig:EW_distribution}.
Therefore, regions of high Galactic latitude are favoured to observe WHIM absorption features, which are known to have large uncertainties on the level of Galactic absorption \citep[e.g.][]{1990ARA&A..28..215D}, and differences between the adopted survey (e.g. \citealp{2005A&A...440..775K} and \citealp{2016A&A...594A.116H}).
However, due to the low levels of Galactic absorption at high latitudes, $\approx$\,1.8\,$\times$\,10$^{20}$\,cm$^{-2}$ for $|b|>60^{\circ}$, it is not foreseen that these large uncertainties will significantly affect the ability to detect absorption features of the WHIM.

The combined line significance required to claim a real detection with high certainty depends on the number of independent spectral elements within the blind line search.
A 4.5$\sigma$ combined detection for 100 independent redshift trials of a blind line search, as in this survey, results in a probability of 0.4$\%$ that the fluctuations could be random.
However, a 4.5$\sigma$ combined significance, or 3.2$\sigma$ each, for 40 spectral elements searched in the spectrum results in a 0.08$\%$ chance of the fluctuations being random.
A reduced blind line search region could be achieved with a low redshift GRB as the background source or with an a priori redshift from UV data or known galaxy locations, as in \citet{2019ApJ...872...83K} and \citet{2020A&A...634A.106A}.

This work assumes a high intrinsic hydrogen column density of 10$^{22}$\,cm$^{-2}$ for the host galaxy of the GRB.
The exact origin of this intrinsic absorption in afterglow spectra has been studied \citep[e.g.][]{2011A&A...525A.113S, 2013ApJ...768...23W}.
\citet{2009MNRAS.397.1177E} present the distribution of intrinsic column density of GRB afterglows observed by \textit{Swift}-XRT, peaking at approximately 10$^{21}$\,cm$^{-2}$, when GRBs of unknown redshift are located at $z$\textsubscript{GRB}\,=\,0, and peaking at 10$^{22}$\,cm$^{-2}$, when $z$\textsubscript{GRB}\,=\,2.23 for those with no recorded redshift value.
A mean value of $\approx$\,5\,$\times$\,10$^{21}$\,cm$^{-2}$ was found by \citet{2012MNRAS.421.1697C} when using only afterglows of known redshift in the sample.
This absorption from the host galaxy of the afterglow will have a similar impact to that of the Milky Way, meaning a lower intrinsic column will increase the significance of these line detections.
Using the \textit{Swift}-XRT population of observed GRB afterglows of known redshift, a relationship has been seen between the intrinsic column density of the host galaxy and the redshift of the GRB where a lower redshift is associated with a lower intrinsic hydrogen column density \citep{2012MNRAS.421.1697C}.
However, it has also been suggested that this observation may be due to a bias in dust extinction \citep{2012ApJ...754...89W}.
If this relationship holds, then targeting lower redshift GRB afterglows will result in a lower intrinsic hydrogen column density.
This can be combined with observing afterglows along lines of sight with a low hydrogen column density in the Milky Way to increase the chance of a significant detection of the WHIM.

\section{Conclusions} \label{sec:conclusions}
This work has assessed the capability of the \textit{Athena} X-IFU to detect the missing baryons of the WHIM, through \ion{O}{vii} and \ion{O}{viii} absorption features at rest-frame energies of 574\,eV and 654\,eV, respectively, in the X-ray afterglow spectra of GRBs. The results of this study are summarised as follows. 
\begin{enumerate}
\item To obtain a 75\% detection probability of \ion{O}{vii} and \ion{O}{viii} with a minimum observed equivalent width of EW\textsubscript{\ion{O}{vii}}$=$\,0.13--0.39\,eV and EW\textsubscript{\ion{O}{viii}}$=$\,0.09--0.26\,eV, \textit{Athena} should target GRB afterglows of \textit{F}\textsubscript{start}\,$\geq$\,2$\,\times\,$10$^{-11}$\,erg\,cm$^{-2}$\,s$^{-1}$, or $\gtrapprox$\,10$^{6}$ high resolution X-IFU counts, in the 0.3--10\,keV energy band. 
This results in 15--30 GRBs\,yr$^{-1}$ within \textit{Athena}'s field of regard, allowing for the detection of 11--34 absorbers per year through \ion{O}{vii}-\ion{O}{viii} pairs with the X-IFU.
\\   
\item Absorption systems weaker than 0.13\,eV will require GRB afterglows of \textit{F}\textsubscript{start}\,>\,10$^{-10}$\,erg\,cm$^{-2}$\,s$^{-1}$, or $\gtrapprox$\,5\,$\times$\,10$^{6}$ high resolution X-IFU counts, to obtain a 75\% chance of detection. This amounts to less than 1\% of the GRB population if the observation begins four hours after the GRB trigger. A shorter ToO reaction time, or longer observation, would be required to observe an afterglow with an observed starting flux greater than 10$^{-10}$\,erg\,cm$^{-2}$\,s$^{-1}$, to reveal absorption systems with observed EW\textsubscript{\ion{O}{vii}}\,<\,0.13\,eV.
\\
\item The X-IFU is capable of measuring the observed equivalent width of narrow absorption features within a 1$\sigma$ limit for features with EW\,$\geq$\,0.14\,eV imprinted on afterglow spectra with \textit{F}\textsubscript{start}\,$\geq$\,5$\,\times\,$10$^{-11}$\,erg\,cm$^{-2}$\,s$^{-1}$. 
\\
\item The absorption from hydrogen in the local Galaxy and the redshift of WHIM absorbers impacts the significance at which absorption features are detected in the X-IFU spectra. The significance of lines decreases with increasing hydrogen column density and with increasing redshift of the WHIM. To obtain a combined line significance of 4.5$\sigma$ of an \ion{O}{vii}-\ion{O}{viii} absorption pair, a hydrogen column density of $N$\textsubscript{H}\,$\leq$\,8\,$\times$\,10$^{20}$\,cm$^{-2}$ along the line of sight is required. This excludes observations of 35\% of the sky, located between an average of 22$^{\circ}$ North and 24$^{\circ}$ South of the Galactic plane. 
\end{enumerate}

The results show that \textit{Athena} will be able to detect the WHIM through \ion{O}{vii}-\ion{O}{viii} absorption features using GRB X-ray afterglows as the background source. 
This study provides limits to the brightness and location of the GRB afterglows that can be observed to reveal the missing baryons of the WHIM with high significance. 
Depending on the triggering mechanism, the detection rate of GRBs, and the distribution of the WHIM, the work presented shows that \textit{Athena} may detect $\approx$\,45--137 absorbers through \ion{O}{vii}-\ion{O}{viii} absorption pairs with EW\textsubscript{\ion{O}{vii}}$\geq$0.13\,eV in GRB afterglows throughout the four-year mission.

\begin{acknowledgements}
The authors thank the anonymous referee for comments that improved the paper. The authors acknowledge useful discussions with Keith Arnaud. SW, SMB, and AMC acknowledge support from the European Space Agency under PRODEX contract number 4000120713. This work was funded in part by the Bundesministerium f\"{u}r Forschung und Technologie under DLR grant 50\,QR\,1903. This work made use of data supplied by the UK Swift Science Data Centre at the University of Leicester. This research made use of Astropy,\footnote{http://www.astropy.org} a community-developed core Python package for Astronomy \citep{astropy:2013, astropy:2018}.
\end{acknowledgements}
\bibliographystyle{aa}
\bibliography{athena}

\begin{appendix}
\section{Predicting detection probability of the WHIM using Poisson statistics}
The probability of detecting unresolved absorption features of Gaussian form within modelled astronomical spectra can be predicted using binomial statistics.
The statistical significance required to detect a line of a given equivalent width EW is calculated by
\begin{equation} \label{nsigma}
\sigma = \dfrac{\mathrm{EW} \cdot \Delta E}{\mathrm{SNRE}},
\end{equation}
where $\Delta E$\,=\,2.5\,eV and SNRE is the signal-to-noise ratio per resolution element, which is calculated from the photon flux across the resolution elements in the energy range of the blind-line search.
The binomial distribution determines the probability of detecting the absorption feature, out of $N$ possible trials, for a one-sided probability associated with the $\sigma$ value calculated using Eq.~\ref{nsigma}.

\begin{figure}
\centering
\resizebox{\hsize}{!}{\includegraphics{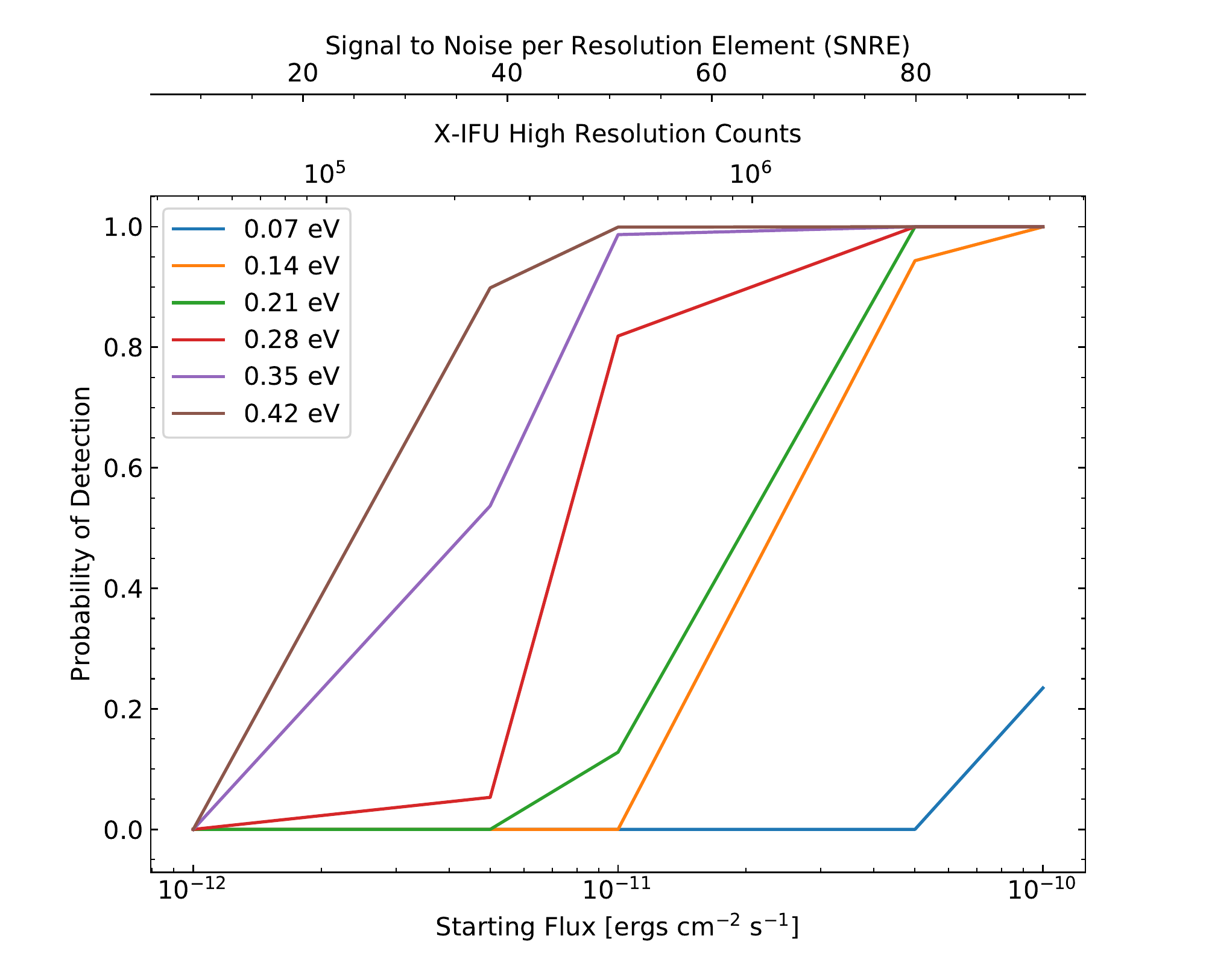}}
\caption{Detection probability of an unresolved absorption feature of a given equivalent width for GRB afterglow starting fluxes of 1, 5, 10, 50, and 100\,$\times$\,10$^{-12}$\,erg\,cm$^{-2}$\,s$^{-1}$.}
\label{fig:theoretical_fluxes}
\end{figure}

Figure~\ref{fig:theoretical_fluxes} shows the probability of detecting an unresolved absorption feature of given equivalent width, in $N$\,=\,250 redshift trials, for GRB afterglow starting fluxes of 1, 5, 10, 50, and 100\,$\times$\,10$^{-12}$\,erg\,cm$^{-2}$\,s$^{-1}$.
The detection probability presented in Sect.~\ref{sec:results} are, on average, approximately a factor of 1.3 less than that of the theoretical results for a given equivalent width and starting flux of a GRB afterglow.
The discrepancy between the detection probability from both methods could arise for a number of reasons.
Firstly, the theoretical method predicts an ideal system in which instrument effects cannot easily be accounted for, such as the detector grading scheme or crosstalk between pixels, we assume a fully linear system, and we do not assume any additional spectral contribution, such as source background events.
Secondly, the spectral binning and fit statistic chosen during the analysis can effect the detection of weak lines. 
Finally, the theoretical results cannot account for the interference of spectral edges on the detection of absorption features depending on their located redshift.

\end{appendix}

\end{document}